\documentclass[%
 aip,
 amsmath,amssymb,
 reprint,%
]{revtex4-1}

\usepackage{graphicx}
\usepackage{dcolumn}
\usepackage{bm}

\usepackage[utf8]{inputenc}
\usepackage[T1]{fontenc}
\usepackage{mathptmx}
\usepackage{etoolbox}

\makeatletter
\def\@email#1#2{%
 \endgroup
 \patchcmd{\titleblock@produce}
  {\frontmatter@RRAPformat}
  {\frontmatter@RRAPformat{\produce@RRAP{*#1\href{mailto:#2}{#2}}}\frontmatter@RRAPformat}
  {}{}
}%
\makeatother
\begin{document}

\preprint{}

\title{Empirical One-Step Conditional Entropy in Infinite Ergodic Systems: Vanishing Entropy Rate, Sparse-Transition Scaling, and Mittag--Leffler Fluctuations}

\author{Ken-ichi Okubo}
\affiliation{
    School of General and Management Studies,
    Suwa University of Science,
    5000-1, Toyohira, Chino, Nagano 391-0292, Japan
}
\email{okubo\_kenichi@rs.sus.ac.jp}

\date{\today}

\begin{abstract}
Empirical entropy rates are widely used to quantify unpredictability
from symbolic or time-series data, yet their interpretation is subtle in
weakly chaotic dynamics, where ordinary Lyapunov exponents vanish and
invariant measures are infinite. We address this issue by studying the
empirical one-step conditional entropy for the fixed finite partitions
considered below in one-dimensional intermittent maps with infinite
invariant measures. For the modified Bernoulli map and the Boole
transformation in the infinite-measure weak-chaos regime, we prove that
this per-step empirical entropy converges to zero. Thus, the usual
entropy-rate normalization becomes asymptotically blind to subexponential
instability. The finite-time information sum, however, remains
informative. Rare transitions between long laminar phases occur on the
return-sequence scale, and their empirical self-information contributes
an additional logarithmic factor. Under the stated regularity and moment
assumptions, this mechanism yields a two-term estimate for the ensemble
mean decay, supported by numerical simulations. Although the raw entropy
rate vanishes, self-normalized fluctuations remain nontrivial and are
numerically consistent with normalized Mittag--Leffler laws. A comparison
with generalized Lyapunov sums shows that the corresponding information
sum is not a Krengel entropy estimator, but a computable,
partition-dependent finite-time measure of sparse symbolic transitions.
These results clarify what empirical Markov entropy can, and cannot,
measure in infinite-measure weak chaos.
\end{abstract}

\maketitle

\begin{quotation}
Empirical entropy and Markov-transition measures are widely used to
quantify unpredictability from observed data. In weakly chaotic systems
with infinite invariant measures, however, ordinary Lyapunov exponents
vanish and trajectories consist of long laminar episodes interrupted by
rare bursts. This paper shows that a familiar one-step empirical entropy
rate becomes asymptotically blind in this setting: for the fixed
partitions considered here, it converges to zero. The vanishing is not a
loss of all symbolic information. When the same data are viewed as a
finite-time information sum, rare transitions carry a logarithmic
self-information contribution and are observed numerically to produce
nontrivial Mittag--Leffler-type normalized fluctuations. Thus, empirical
Markov entropy should be interpreted not as a universal entropy invariant
but as a computable finite-time probe of sparse transitions in weak
chaos.
\end{quotation}

\section{Introduction}
The Lyapunov exponent is a fundamental tool for characterizing dynamical instability.  In many finite-measure chaotic systems, a positive Lyapunov exponent accompanies an exponential separation of trajectories and a positive Kolmogorov--Sinai entropy.  However, complex dynamics also occur in systems where the ordinary Lyapunov exponent is zero.  Examples include billiard systems \cite{zaslavsky2007physics}, random dynamical systems \cite{lai2003noise,wang2004strange}, strange nonchaotic attractors \cite{grebogi1984strange}, and infinite ergodic systems with subexponential instability \cite{pomeau1980intermittent,gaspard1988sporadicity}.  In the latter class, long laminar phases near indifferent fixed points lead to intermittency.

Infinite ergodic systems preserve invariant measures that are not normalizable.  They have been studied in terms of their intermittency \cite{pomeau1980intermittent}, nonstationary chaos \cite{shinkai2006lempel,klages2013weak}, weak ergodicity breaking \cite{bel2006weak}, anomalous diffusion \cite{geisel1984anomalous,geisel1985accelerated}, aging \cite{barkai2003aging,akimoto2013aging}, and laser cooling processes \cite{barkai2021transitions,barkai2022gas}.  In finite invariant probability spaces, Birkhoff averages of integrable observables converge to the phase averages.  In infinite-measure dynamical systems, suitably scaled time averages of observables satisfying certain conditions converge in distribution, as in the Darling--Kac--Aaronson framework.

Several weak-chaos indicators have been developed for such systems.  These include generalized Lyapunov exponents \cite{akimoto2003logarithmic,korabel2009pesin,korabel2010separation,akimoto2015generalized}, Lyapunov pairs \cite{akimoto2010subexponential}, and the Lempel--Ziv complexity \cite{shinkai2006lempel,shinkai2007lempel,korabel2010separation}.  In particular, Korabel and Barkai related the generalized Lyapunov exponent to Krengel entropy through a generalized Pesin identity \cite{korabel2009pesin,korabel2010separation}.  These results motivate a related but distinct question: What information-theoretic content is captured by a simple empirical transition model constructed directly from a symbolic time series?

The quantity considered here is the empirical one-step conditional entropy of a fixed finite partition.  It is built from the empirical state and transition frequencies and is commonly interpreted, in finite-measure settings, as a finite-data one-step Markov approximation to an entropy rate or Kolmogorov--Sinai-type entropy \cite{cohen1985computing,arnold1968ergodic,gaspard1993noise,lesne2014shannon}.    Related transition-model quantities also appear in the literature under names such as the ``entropic chaos degree'' \cite{ohya1998complexities,inoue2000application,inoue2021improved,inoue2022analysis,inoue2023quantification,mao2019investigation,mao2026chaotic}.  We use the descriptive term ``empirical one-step conditional entropy'' to emphasize the concrete finite-time object studied in this paper.

Our first result is a negative theorem for the usual per-step normalization.  For fixed finite partitions of the modified Bernoulli map and the Boole transformation, the empirical one-step conditional entropy \(H_P^{(1,n)}\) converges to zero in an infinite-measure regime.  However, this does not mean that symbolic dynamics contain no information.  Rather, the orbit spends most of its time in long laminar phases such that
the empirical one-step transition matrix becomes nearly diagonal and is
dominated by self-transition.  Consequently, the average one-step information per unit time degenerates.

The second part of the paper estimates the rate at which this vanishing occurs in the ensemble mean.  For the dyadic partition, transitions between the two laminar regions occur only during rare escapes from neighborhoods of indifferent fixed points.  Their expected empirical frequency is of the order \(a_n/n\), where \(a_n\) is the return sequence.  By combining this sparse-transition estimate with the small-probability expansion
of binary entropy under regularity assumptions for non-leading
transition-type terms and first/logarithmic-moment control of the normalized
transition count, we obtain
\[
    \left\langle H_P^{(1,n)}\right\rangle_{\rm ens}
    \simeq
    c_{1,P}\frac{a_n}{n}\log\frac{n}{a_n}
    +c_{2,P}\frac{a_n}{n}.
\]
Similarly, the information sum \(I_P(n)=nH_P^{(1,n)}\) contains a sparse-transition count weighted by a logarithmic self-information factor.  Numerical simulations of the modified Bernoulli map and the Boole transformation support this two-term order estimate and show that the second term is visible at accessible times, particularly near the transition to the infinite-measure regime.  The logarithmic factor distinguishes the present empirical Markov information sum from the generalized Lyapunov sums and Lempel--Ziv complexity, which belong to the return-sequence scale \(a_n\).

Although the raw entropy rate vanishes, its trajectory-to-trajectory fluctuations remain informative after self-normalization.  The Darling--Kac--Aaronson theorem suggests Mittag--Leffler-type fluctuations for Birkhoff sums of fixed \(L^1(\mu)\) observables.  However, the observable defining \(I_P(n)\) depends on the empirical transition matrix and, hence, on the observation time \(n\); therefore, the theorem does not apply directly.  Nevertheless, the sparse-transition picture suggests that the dominant randomness of \(I_P(n)\) is inherited from the same intermittent return activity.  This motivates the self-normalized variable
\[
    Z_P^{(n)}
    =
    \frac{H_P^{(1,n)}}{\langle H_P^{(1,n)}\rangle_{\rm ens}}
    =
    \frac{I_P(n)}{\langle I_P(n)\rangle_{\rm ens}}.
\]
Any deterministic normalization, including the leading scale \(I_P/(a_n\log(n/a_n))\), cancels in this ratio.  Therefore, the purpose of this normalization is not to define a new absolute entropy rate but to isolate the residual weak-chaos fluctuations after the deterministic mean decay has been removed.

Finally, we compare \(I_P(n)\) with the finite-time generalized Lyapunov sum
\[
    G_n=\sum_{k=0}^{n-1}\log|T'(x_k)|.
\]
Because \(G_n\) is a Birkhoff sum of a fixed integrable observable, it belongs to the return-sequence scale.  In contrast, \(I_P(n)\) contains the extra rare-transition
self-information factor \(\log(n/a_n)\).  Thus, \(I_P(n)\) should not be interpreted as a Krengel-entropy estimator or a new generalized Pesin identity.  However, numerically, the ensemble ratio \(\langle I_P(n)\rangle_{\rm ens}/\langle G_n\rangle_{\rm ens}\) is well described by a linear function of \(L_n=\log(n/a_n)\) with a partition-dependent slope and intercept.  This provides a finite-time relationship between the empirical Markov information sum and the intermittent activity measured by \(G_n\).

Unless otherwise stated, the ensemble averages in the numerical sections are sample averages over independent initial conditions drawn from the Lebesgue distribution in the unit interval.  The partition is always fixed as \(n\) varies.  This fixed-partition setting is essential: The coefficients appearing below depend on the partition and should not be interpreted as universal entropy invariants.

We also emphasize the operational characteristics of the proposed
quantity.  The primary numerical tests use fixed cylinder
partitions to maintain symbolic observation precisely
controlled.  However, the construction of \(H_P^{(1,n)}\)
requires only a symbolic time series and empirical
transition matrix.  To illustrate this, Appendix~\ref{app:equal-bin}
presents a map-blind equal-bin robustness check in which the
partition is constructed directly from the observed scalar
time series without an explicit map form, its
derivative, an infinite invariant density, or a cylinder partition.  The resulting self-normalized distributions remain
consistent with the normalized Mittag--Leffler laws for the
benchmark maps considered here.

The remainder of this paper is organized as follows.  In Sec.~II, we define the empirical one-step conditional entropy from the finite-time transition probabilities.  In Sec.~III, we prove its vanishing for the modified Bernoulli map and the Boole transformation.  In Sec.~IV, we develop a sparse-transition scaling estimate for the ensemble mean and identify the assumptions behind the two-term form.  In Sec.~V, we study the self-normalized fluctuation variable and compare it with normalized Mittag--Leffler laws.  In Sec.~VI, we relate the information sum to generalized Lyapunov sums and discuss the resulting partition-dependent relationship between the empirical Markov information sum and intermittent activity.  Finally, Sec.~VII summarizes the implications, limitations, and possible extensions of the framework.

\section{One-step conditional entropy from empirical transition probabilities}
In this section, we first recall the stationary one-step conditional entropy for finite-measure systems as a reference and then define the empirical finite-time quantity used throughout the remainder of the paper. The latter requires only a symbolic time series and does not assume the existence of a normalizable invariant measure.
The finite-length expression given below is used in subsequent sections to analyze weak chaos in infinite-measure systems.

Let $(X,T,\mu)$ be a measure-preserving dynamical system, where $X$ is the phase space, $T:X\to X$ is a map, and $\mu$ is an invariant probability measure. Let $P=\{X_i\}_{i=1}^N$ be a finite partition of $X$. For an initial point $x_0\in X$, the orbit $\{x_k\}_{k\ge 0}$ generated by $x_k=T^k(x_0)$ induces a symbolic sequence $\{s_k\}_{k\ge 0}$ defined by
\begin{align*}
    s_k = i \Leftrightarrow x_k \in X_i.
\end{align*}

When the system is in a stationary state, the state, joint, and one-step transition probabilities associated with the partition are given by
\begin{align}
    \rho_i &\equiv \mu(X_i)=\mathbb{P}(s_k=i)\notag,\\
    r_{ij} &\equiv \mu(X_i\cap T^{-1}(X_j))=\mathbb{P}(s_k=i,s_{k+1}=j)\notag,\\
    t_{ij} &\equiv \frac{r_{ij}}{\rho_i}=\frac{\mathbb{P}(s_k=i,s_{k+1}=j)}{\mathbb{P}(s_k=i)}=\mathbb{P}(s_{k+1}=j|s_k=i)~ (\rho_i \neq 0)\notag.
\end{align}
These quantities satisfy
\begin{align}
    r_{ij} = t_{ij}\rho_i,~\sum_{j}r_{ij}=\rho_i,~\sum_j t_{ij} &= 1 ~(\rho_i\neq 0).\notag
\end{align}

We define the one-step conditional entropy of the partition-induced symbolic process as
\begin{align}
    H_P^{(1)} = -\sum_{i=1}^N \sum_{j=1}^N \rho_i t_{ij}\log t_{ij}.
\end{align}
This coincides with the conditional entropy $H_P(s_{k+1}\mid s_k)$ of the symbolic process.

Next, from a finite-length time series $\{s_k\}_{0\le k\le n}$, we define the empirical counterparts
\begin{equation}
	\rho_i^{(n)}(x_0) \equiv \frac{\#\{k|s_k=i,0\leq k\leq n-1\}}{n}. \label{Def: rho_i^n}
\end{equation}
\begin{equation}
	r_{ij}^{(n)}(x_0) \equiv \frac{\#\{k|(s_k, s_{k+1})=(i,j),0\leq k\leq n-1\}}{n}.
	\label{Def:r_ij^n}
\end{equation}
\begin{equation}
    t_{ij}^{(n)}(x_0) \equiv \frac{r_{ij}^{(n)}(x_0)}{\rho_i^{(n)}(x_0)}, ~(\rho_i^{(n)}(x_0)\neq 0) \label{Def: t_ij^n}.
\end{equation}
If $\rho_i^{(n)}(x_0)=0$, we define $r_{ij}^{(n)}(x_0)=0$ and $t_{ij}^{(n)}(x_0)=0$.
Equivalently, the symbolic sequence must be known for up to $s_n$:
$\rho_i^{(n)}(x_0)$ is computed from $s_0,\ldots,s_{n-1}$,
whereas $r_{ij}^{(n)}(x_0)$ is computed from the transition pairs
$(s_k,s_{k+1})$, $0\leq k\leq n-1$.  If each symbol $s_k$
is determined directly from the phase-space point $x_k$, this corresponds
to computing the scalar orbit up to $x_n$.  However, for a depth-$m$ dyadic
cylinder partition, the state at time $k$ is represented by the
word
\[
    s_k^{(m)}=(b_k,b_{k+1},\ldots,b_{k+m-1}),
\]
where $b_k$ is the dyadic symbol of $x_k$.  Here, $n$ denotes the
number of empirical cylinder-state transitions: the occupation counts
$\rho_i^{(n)}(x_0)$ require an underlying scalar orbit up to
$x_{n+m-2}$, whereas the transition counts $r_{ij}^{(n)}(x_0)$ require it
up to $x_{n+m-1}$.  Because $m$ is fixed, this finite offset does not affect
the asymptotic normalization by $n$.

Using these empirical quantities, we define the one-step conditional entropy of the empirical transition model as
\begin{align}
    H_P^{(1,n)}(x_0) = -\sum_{i=1}^N \sum_{j=1}^N \rho_i^{(n)}(x_0) t_{ij}^{(n)}(x_0) \log t_{ij}^{(n)}(x_0)\label{Eq: ECD finite length 2},
\end{align}
where if $t_{ij}^{(n)}(x_0)=0$, we define $t_{ij}^{(n)}(x_0)\log t_{ij}^{(n)}(x_0)=0$.
Equivalently,
\begin{align}
    H_P^{(1,n)}(x_0)=\sum_{i=1}^N \sum_{j=1}^N r_{ij}^{(n)}(x_0) \log \frac{\rho_i^{(n)}(x_0)}{r_{ij}^{(n)}(x_0)} \label{Eq: ECD finite length},
\end{align}
where if $r_{ij}^{(n)}(x_0)=0$, we define $r_{ij}^{(n)}(x_0) \log \frac{\rho_i^{(n)}(x_0)}{r_{ij}^{(n)}(x_0)}=0$.

Thus, $H_P^{(1,n)}(x_0)$ is an empirical approximation of the one-step conditional entropy of the partition-induced symbolic process. The quantity $H_P(s_{k+1}\mid s_k)$ represents the average uncertainty in predicting the next symbol when the current symbol is known. It is also a one-step Markov approximation of the entropy rate and is related, for suitable finite partitions and refinements, to data-based approximations of Kolmogorov--Sinai entropy \cite{cohen1985computing,gaspard1993noise,lesne2014shannon,arnold1968ergodic}.  

In the following, we show that $H_P^{(1,n)}(x_0)$ can be transformed into the form of a conditional information sum. That is,
\begin{equation}
    H_P^{(1,n)}(x_0) = \frac{1}{n}\sum_{k=0}^{n-1}\phi_P^{(n)}(x_k),
\end{equation}
where $\phi_P^{(n)}(x_k)\equiv -\log t_{s_k s_{k+1}}^{(n)}(x_0)$ is the one-step conditional self-information.

Let
\begin{align}
    N_i^{(n)}(x_0) &\equiv \#\{k|s_k=i, 0\leq k \leq n-1\},\label{Def: N_i^n}\\
    N_{ij}^{(n)}(x_0) & \equiv \#\{k|(s_k,s_{k+1})=(i,j),0\leq k \leq n-1\} \label{Def: N_ij^n}.
\end{align}
Because $\rho_i^{(n)}=N_i^{(n)}/n$ and $r_{ij}^{(n)}=N_{ij}^{(n)}/n$, Eq. \eqref{Eq: ECD finite length} can be written as
\begin{align}
    H_P^{(1,n)}(x_0) &= -\frac{1}{n}\sum_{i=1}^N \sum_{j=1}^N N_{ij}^{(n)}(x_0) \log t_{ij}^{(n)}(x_0)\label{Eq: ECD finite length 3}.
\end{align}

Now, let $g:\{1,\dots,N\}^2\to\mathbb R$ be any function. Subsequently,
\begin{equation}
    \sum_{i=1}^N\sum_{j=1}^N N_{ij}^{(n)}(x_0)g(i,j) = \sum_{k=0}^{n-1}g(s_k, s_{k+1}).
    \label{Eq: (i,j) to (s_k, s_{k+1})}
\end{equation}
Indeed, by definition,
\begin{equation}
    N_{ij}^{(n)}(x_0) = \sum_{k=0}^{n-1}\bm{1}_{\{(i,j)\}}(s_k, s_{k+1}), \label{Eq: N_ij characteristic function}
\end{equation}
such that
\begin{widetext}
\begin{equation}
    \sum_{i,j}N_{ij}^{(n)}g(i,j)=\sum_{i,j}\sum_{k=0}^{n-1}\bm{1}_{\{(i,j)\}}(s_k,s_{k+1})g(i,j)=
    \sum_{k=0}^{n-1}\sum_{i,j}\bm{1}_{\{(i,j)\}}(s_k,s_{k+1})g(i,j).
    \label{Eq: sum_(i,j)N_ij g(i,j)=sum_(k=0)sum_(i,j)1(s_k,s_k+1)g(i,j)}
\end{equation}
\end{widetext}
For a fixed $k$, if $(s_k,s_{k+1})=(a,b)$, then only the pair $(i,j)=(a,b)$ contributes. Therefore,
\begin{equation}
    \sum_{i,j}\bm{1}_{\{(i,j)\}}(s_k,s_{k+1})g(i,j)=g(a,b)=g(s_k, s_{k+1}).
    \label{Eq: sum_(i,j)1(s_k,s_k+1)g(i,j)=g(s_k,s_k+1)}
\end{equation}
Substituting Eq. \eqref{Eq: sum_(i,j)1(s_k,s_k+1)g(i,j)=g(s_k,s_k+1)} into Eq. \eqref{Eq: sum_(i,j)N_ij g(i,j)=sum_(k=0)sum_(i,j)1(s_k,s_k+1)g(i,j)} gives Eq. \eqref{Eq: (i,j) to (s_k, s_{k+1})}.

Selecting $g(i,j)=-\log t_{ij}^{(n)}(x_0)$, we obtain
\begin{align}
    H_P^{(1,n)}(x_0) &= -\frac{1}{n}\sum_{i=1}^N \sum_{j=1}^N N_{ij}^{(n)}(x_0) \log t_{ij}^{(n)}(x_0)\notag\\
    &= \frac{1}{n}\sum_{i,j}N_{ij}^{(n)}(x_0)\left(-\log t_{ij}^{(n)}(x_0)\right)\notag\\
    &= \frac{1}{n}\sum_{k=0}^{n-1}\left(-\log t_{s_k s_{k+1}}^{(n)}\right)\notag\\
    &= \frac{1}{n}\sum_{k=0}^{n-1}\phi_P^{(n)}(x_k) \label{Eq: information sum}
\end{align}

This representation is used in Secs. V and VI. However, note that $\phi_P^{(n)}$ depends on the time horizon $n$ through the empirical transition matrix $t_{ij}^{(n)}$; therefore, it is not a fixed observable in the sense of the standard Darling--Kac--Aaronson theorem.

\section{Vanishing asymptotics of the one-step conditional entropy in infinite ergodic systems}
\begin{figure}[h]
	\centering
	\includegraphics[width=.7\columnwidth]{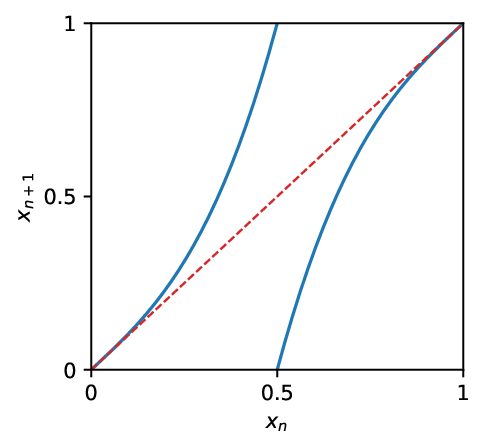}
	\caption{Return map of the modified Bernoulli map for $B_{MB}=3$ expressed by Eq. (\ref{Eq: modified Bernoulli map}).}
	\label{Fig: modified Bernoulli map}
\end{figure}

\begin{figure}[h]
	\centering
	\includegraphics[width=.7\columnwidth]{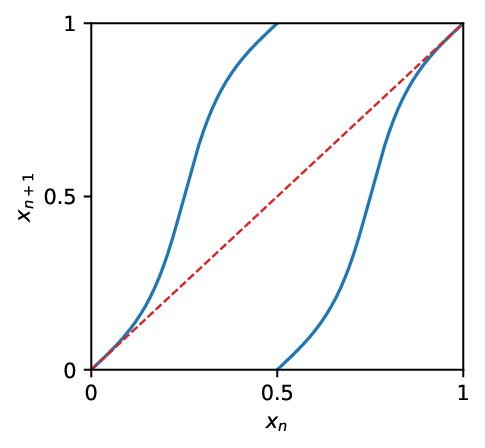}
	\caption{Return map of the Boole transformation expressed by Eq. (\ref{Eq: Boole modified}).}
	\label{Fig: Boole transformation}
\end{figure}

In this section, we derive the theoretical asymptotic behavior of the one-step conditional entropy for two representative infinite ergodic maps: the modified Bernoulli map \cite{aizawa1989stagnant} and the Boole transformation \cite{adler1973ergodic}. 
The modified Bernoulli map is given by
\begin{widetext}
\begin{equation}
		x_{n+1} = T_{B_{MB}}(x_n)= \left\lbrace
		\begin{array}{ll}
			x_n + 2^{B_{MB}-1}x_n^{B_{MB}}, & x_n \in I_0 = [0,1/2)\\
			x_n-2^{B_{MB}-1}(1-x_n)^{B_{MB}}, & x_n \in I_1 = [1/2,1],
		\end{array}
		\right.
	\label{Eq: modified Bernoulli map}
\end{equation}
\end{widetext}
and the Boole transformation by 
\begin{equation}
	y_{n+1} = T(y_n) = y_n -\frac{1}{y_n},~y_n \in \mathbb{R}.
\end{equation}
For the Boole transformation, we use the standard variable transformation $y_n=-\cot(\pi x_n)$ \cite{umeno2016exact}, which yields
\begin{equation}
	x_{n+1} =T(x_n)= \frac{1}{\pi}\cot^{-1}\left\lbrace2\cot(2\pi x_n)\right\rbrace. \label{Eq: Boole modified}
\end{equation}
Figures \ref{Fig: modified Bernoulli map} and \ref{Fig: Boole transformation} show the return maps of the modified Bernoulli map and the Boole transformation, respectively.

The modified Bernoulli map for $B_{MB}>2$ and the Boole transformation preserve an infinite ergodic measure.
The modified Bernoulli map is used to study non-equilibrium non-stationary physics \cite{akimoto2007new,akimoto2005large,akimoto2003logarithmic}. 
The Boole transformation has been studied from the perspective of the route of chaos \cite{umeno2016exact,okubo2018universality,okubo2022universal}, and extensions to the Boole transformation have been studied intensively \cite{okubo2021infinite,prykarpatski2024two}.

In this section, we focus directly on the empirical quantities introduced in Sec. II rather than assuming the existence of pointwise infinite-time limits of $\rho_i^{(n)}$, $r_{ij}^{(n)}$, or $t_{ij}^{(n)}$. 

We assume that the partition $P=\{X_i\}_{i=1}^{N}$ is either an equipartition or a cylinder partition constructed as an $m$-step refinement of the dyadic partition ${[0,1/2),[1/2,1]}$, and $X_1$ and $X_N$ contain indifferent fixed points at the two ends of the interval.

According to a corollary of Hopf's ergodic theorem for infinite ergodic measures \cite{aaronson1997introduction}, if $A$ is a measurable set with a finite invariant measure, its occupation frequency converges to zero for almost every initial condition. In particular,
\begin{equation}
		\lim_{n\to\infty}\frac{\#\{k|x_k\in A,0\leq k\leq n-1\}}{n}=0~~\mbox{for almost every}~x_0.
        \label{Eq: Hopf's ergodic theorem}
\end{equation}
Applying Eq. \eqref{Eq: Hopf's ergodic theorem} to each interior cell $X_i$ with $i\neq 1,N$, we obtain
\begin{equation*}
    \rho_i^{(n)}(x_0) \to 0~~\mbox{for almost every}~x_0,~~~(i\neq 1, N).
\end{equation*}
Hence, the empirical occupation frequency of any interior partition element asymptotically vanishes.

Next, according to Thaler's limit theorem for sojourns near indifferent fixed points \cite{thaler2002limit}, for the modified Bernoulli map and the Boole transformation, and for any neighborhood $U=[0,\gamma]$ with $0<\gamma<1$, we obtain
\begin{equation}
	\frac{\#\{k|x_k\in U,0\leq k\leq n-1\}}{n} \overset{d}{\to} L_{\alpha,\beta}~~(n\to \infty),
	\label{Eq: Thaler}
\end{equation}
where $L_{\alpha,\beta}$ is a random variable with a generalized arcsine distribution. For the modified Bernoulli map and the Boole transformation, $\beta=1/2$; therefore, $L_{\alpha,\beta}$ and $1-L_{\alpha,\beta}$ have the same distribution.

Therefore, for the empirical occupation frequencies, we obtain
\begin{equation}
	\rho_i^{(n)}(x_0) \overset{d}{\to} \left\lbrace
	\begin{array}{lll}
		L_{\alpha,\beta},& i=1,\\
		1-L_{\alpha,\beta},& i=N,\\
		0,& \mbox{otherwise}.
	\end{array}
	\right.
	\label{Eq:rho_i}
\end{equation}
Because Eq. \eqref{Eq: Thaler} holds for any sufficiently small neighborhood of an indifferent fixed point, it applies in particular to $X_1\cap T^{-1}(X_1)$ and $X_N\cap T^{-1}(X_N)$. Hence,
\begin{equation}
	r_{ij}^{(n)}(x_0) \overset{d}{\to} \left\lbrace
	\begin{array}{lll}
		L_{\alpha,\beta},& (i,j)=(1,1),\\
		1-L_{\alpha,\beta},& (i,j)=(N,N),\\
		0,& \mbox{otherwise}. 
	\end{array}
	\right.
	\label{Eq:r_ij}
\end{equation}
More precisely, for $(i,j)\neq(1,1),(N, N)$, the set $X_i\cap T^{-1}(X_j)$ has a finite invariant measure and does not contain an indifferent fixed point. Thus, Eq. \eqref{Eq: Hopf's ergodic theorem} yields
\begin{equation}
    r_{ij}^{(n)}(x_0) \to 0~~\mbox{for almost every}~x_0,~~((i,j)\neq(1,1), (N,N)). 
    \label{Eq: r_ij neq 11 NN}
\end{equation}

In particular, we define
\begin{equation*}
    \delta_1^{(n)}(x_0) \equiv \rho_1^{(n)}(x_0)-r_{11}^{(n)}(x_0),~~~\delta_N^{(n)}(x_0)\equiv \rho_N^{(n)}(x_0)-r_{NN}^{(n)}(x_0).
\end{equation*}

Because
\begin{equation*}
    \delta_1^{(n)}(x_0)=\sum_{j\neq 1}r_{1j}^{(n)}(x_0),~~~\delta_N^{(n)}(x_0)=\sum_{j\neq N}r_{Nj}^{(n)}(x_0),
\end{equation*}
Eq. \eqref{Eq: r_ij neq 11 NN} implies
\begin{equation}
    \delta_1^{(n)}(x_0)\to 0,~~~\delta_N^{(n)}(x_0)\to 0~~\mbox{for almost every} ~x_0.
    \label{Eq: delta 1,N converge to 0}
\end{equation}
We now evaluate the one-step conditional entropy using Eq. \eqref{Eq: ECD finite length}. First, we consider the terms with $(i,j)\neq(1,1),(N,N)$. As
\begin{equation*}
    0\leq r_{ij}^{(n)}\leq \rho_i^{(n)}(x_0)\leq 1,
\end{equation*}
we obtain
\begin{equation*}
    0\leq r_{ij}^{(n)}(x_0)\log \frac{\rho_i^{(n)}(x_0)}{r_{ij}^{(n)}(x_0)}
    \leq r_{ij}^{(n)}(x_0) \log \frac{1}{r_{ij}^{(n)}(x_0)}.
\end{equation*}
Because $r_{ij}^{(n)}(x_0)\to 0$ almost everywhere by Eq. \eqref{Eq: r_ij neq 11 NN}, and because $x\log(1/x)\to 0$ as $x\to +0$, we obtain
\begin{equation}
    r_{ij}^{(n)}\log \frac{\rho_i^{(n)}(x_0)}{r_{ij}^{(n)}(x_0)}\to 0~~\mbox{for almost every}~ x_0
    ~((i,j)\neq (1,1), (N,N)).
    \label{Eq: rlog rho/r converge to 0}
\end{equation}
Next, we consider the diagonal boundary terms. Because
\begin{equation*}
    \rho_1^{(n)}(x_0) = r_{11}^{(n)}(x_0) + \delta_1^{(n)}(x_0),
\end{equation*}
we can write
\begin{equation*}
    r_{11}^{(n)}(x_0)\log \frac{\rho_1^{(n)}(x_0)}{r_{11}^{(n)}(x_0)}=
    r_{11}^{(n)}(x_0)\log \left(1+\frac{\delta_1^{(n)}(x_0)}{r_{11}^{(n)}(x_0)}\right).
\end{equation*}
Using the inequality $\log(1+u)\le u$ for $u\ge 0$, we obtain
\begin{equation*}
    0\le r_{11}^{(n)}(x_0)\log \frac{\rho_1^{(n)}(x_0)}{r_{11}^{(n)}(x_0)}\le \delta_1^{(n)}(x_0),
\end{equation*}
which converges to zero almost everywhere by Eq. \eqref{Eq: delta 1,N converge to 0}. The same argument applies to the right boundary term involving $r_{NN}^{(n)}(x_0)$ and $\delta_N^{(n)}(x_0)$.

By combining these bounds with Eq. \eqref{Eq: rlog rho/r converge to 0}, we conclude that every term in Eq. \eqref{Eq: ECD finite length} tends to zero for almost every initial condition. Therefore,
\begin{equation}
    H_P^{(1,n)}(x_0) \to 0~\mbox{for almost every}~x_0~~(n\to \infty).
    \label{Eq: ECD infinite ergodic infinite length}
\end{equation}
This vanishing limit should not be interpreted as an absence of dynamical complexity. Rather, it shows that in these weakly chaotic infinite-ergodic maps, a fixed-partition empirical one-step conditional entropy loses asymptotic sensitivity under standard per-step normalization: The amount of new information per step associated with the partition becomes asymptotically degenerate. Thus, the standard $1/n$-normalized quantity is asymptotically blind to weak chaos.

This result should be read as a negative theorem for the usual normalization: fixed-partition empirical one-step conditional entropy is not an adequate stand-alone infinite-time descriptor of weak chaos under $1/n$ scaling.  Therefore, this motivates two complementary finite-time questions: how rapidly the ensemble mean decays, and whether the trajectory-to-trajectory fluctuations remain nontrivial after self-normalization.

\section{Scaling estimate for the ensemble mean of the one-step conditional entropy}
\label{sec:mean_scaling}

In the preceding section, we have proved that \(H_P^{(1,n)}(x_0)\to0\) for the
modified Bernoulli map and the Boole transformation.  We now estimate the
decay of its ensemble mean and, equivalently, the growth of the information
sum
\[
    I_P(n,x_0)=nH_P^{(1,n)}(x_0).
\]
The basic mechanism is the sparsity of symbolic transitions: Rare transitions between long
laminar phases occur on the return-sequence scale, and the self-information of such rare transitions provides an additional logarithmic factor.

First, we consider the dyadic partition
\begin{equation*}
  P_2=\{X_1,X_2\},\qquad
  X_1 = \left[0,\frac{1}{2}\right), \qquad
  X_2 = \left[\frac{1}{2},1\right],
\end{equation*}
and define
\begin{equation*}
  \sigma(x) =
  \begin{cases}
    +1, & x\in X_1,\\
    -1, & x\in X_2.
  \end{cases}
\end{equation*}
For a trajectory $\{x_k\}$, let
\begin{equation}
	C(m,n)\equiv \frac{1}{n}\sum_{k=0}^{n-1}\sigma(x_k)\sigma(x_{k+m})
	\label{eq:correlation_function}
\end{equation}
be the finite-time correlation function \cite{akimoto2007new}.  For \(m=1\), we define
\begin{widetext}
\begin{align*}
	A_n(x_0)
	&\equiv
	\#\{k|
    (s_k, s_{k+1}) = (1,1)
	\ {\rm or}\ 
	(s_k, s_{k+1}) = (2,2), 0\le k\le n-1\},\\
	B_n(x_0)
	&\equiv
	\#\{k|
	(s_k, s_{k+1}) = (1,2)
	\ {\rm or}\ 
	(s_k, s_{k+1}) = (2,1),0\le k\le n-1\}.
\end{align*}
\end{widetext}
Subsequently, \(A_n+B_n=n\) and \(A_n-B_n=nC(1,n)\). Hence,
\begin{equation}
    p_n(x_0)
    \equiv
    r_{12}^{(n)}(x_0)+r_{21}^{(n)}(x_0)
    =
    \frac{B_n(x_0)}{n}
    =
    \frac{1-C(1,n)}{2}.
    \label{eq:pn_cross_frequency}
\end{equation}
Thus, \(p_n\) is exactly the empirical frequency of crossings between the two
halves.  Because the transitions \(X_1\to X_2\) and \(X_2\to X_1\) must alternate
in a two-symbol sequence,
\begin{equation}
    r_{12}^{(n)}(x_0)
    =
    \frac{p_n(x_0)}{2}
    +O\!\left(\frac{1}{n}\right),
    \qquad
    r_{21}^{(n)}(x_0)
    =
    \frac{p_n(x_0)}{2}
    +O\!\left(\frac{1}{n}\right).
    \label{eq:r12_r21_pn}
\end{equation}

The empirical one-step conditional entropy for this dyadic partition can be
expressed as
\begin{equation}
    H_{P_2}^{(1,n)}(x_0)
    =
    \rho_1^{(n)}(x_0)
    h\!\left(t_{12}^{(n)}(x_0)\right)
    +
    \rho_2^{(n)}(x_0)
    h\!\left(t_{21}^{(n)}(x_0)\right),
    \label{eq:H_binary_form}
\end{equation}
where
\[
    h(u)=-(1-u)\log(1-u)-u\log u
\]
is the binary entropy function.  For a small \(u\),
\begin{equation}
    h(u)=u\log\frac{1}{u}+u+O(u^2).
    \label{eq:binary_entropy_expansion}
\end{equation}

Now, we relate the small-
\(u\) expansion in Eq.~\eqref{eq:binary_entropy_expansion} to the total
crossing frequency \(p_n\).  The small variables in Eq.~\eqref{eq:binary_entropy_expansion}
are \(t_{12}^{(n)}\) and \(t_{21}^{(n)}\), not \(p_n\) itself.  We first use the
following elementary identity: for \(0\le r\le \rho\),
\begin{equation*}
    \rho h\!\left(\frac{r}{\rho}\right)
    =
    r\log\frac{\rho}{r}
    +
    (\rho-r)\log\frac{\rho}{\rho-r},
\end{equation*}
where the terms with a zero prefactor are understood by continuity.  Therefore,
Eq.~\eqref{eq:H_binary_form} can be expressed as
\begin{align*}
    H_{P_2}^{(1,n)}(x_0)
    =&\;
    r_{12}^{(n)}\log\frac{\rho_1^{(n)}}{r_{12}^{(n)}}
    +
    r_{21}^{(n)}\log\frac{\rho_2^{(n)}}{r_{21}^{(n)}}
    \\
    &+
    \left(\rho_1^{(n)}-r_{12}^{(n)}\right)
    \log\frac{\rho_1^{(n)}}{\rho_1^{(n)}-r_{12}^{(n)}}
    \\
    &+
    \left(\rho_2^{(n)}-r_{21}^{(n)}\right)
    \log\frac{\rho_2^{(n)}}{\rho_2^{(n)}-r_{21}^{(n)}}.
\end{align*}
The first two terms represent the contributions of cross-transitions, whereas the last
two terms are the contributions of self-transitions.

First, we estimate the self-transition contribution.  For \(0\le r\le \rho\),
\begin{equation*}
    0\le
    (\rho-r)\log\frac{\rho}{\rho-r}
    \le r.
\end{equation*}
This follows from \(\log x\le x-1\) with \(x=\rho/(\rho-r)\).  Hence, the
self-transition contribution is bounded by
\begin{equation*}
    O\!\left(r_{12}^{(n)}+r_{21}^{(n)}\right)=O(p_n).
\end{equation*}

Next, we consider the cross-transition contribution.  We define
\begin{equation*}
    q_{1,n}=r_{12}^{(n)},\qquad
    q_{2,n}=r_{21}^{(n)},\qquad
    p_n=q_{1,n}+q_{2,n}.
\end{equation*}
If \(p_n>0\), we write
\begin{equation*}
    q_{i,n}=p_n\theta_{i,n},
    \qquad
    \theta_{1,n}+\theta_{2,n}=1.
\end{equation*}
Here \(\theta_{1,n}\) is the fraction of cross-transitions of type \(X_1\to X_2\),
and \(\theta_{2,n}\) is the fraction of cross transitions of type \(X_2\to X_1\).
Subsequently,
\begin{align*}
    &r_{12}^{(n)}\log\frac{\rho_1^{(n)}}{r_{12}^{(n)}}
    +
    r_{21}^{(n)}\log\frac{\rho_2^{(n)}}{r_{21}^{(n)}}
    \\
    &\quad=
    p_n\theta_{1,n}
    \log\frac{\rho_1^{(n)}}{p_n\theta_{1,n}}
    +
    p_n\theta_{2,n}
    \log\frac{\rho_2^{(n)}}{p_n\theta_{2,n}}
    \\
    &\quad=
    p_n\theta_{1,n}
    \left(\log\frac{1}{p_n}+\log\frac{\rho_1^{(n)}}{\theta_{1,n}}\right)
    +
    p_n\theta_{2,n}
    \left(\log\frac{1}{p_n}+\log\frac{\rho_2^{(n)}}{\theta_{2,n}}\right)
    \\
    &\quad=
    p_n\log\frac{1}{p_n}
    +
    p_n\sum_{i=1}^{2}\theta_{i,n}
    \log\frac{\rho_i^{(n)}}{\theta_{i,n}}.
\end{align*}
Thus, the leading logarithmic contribution is governed by
\(p_n\log(1/p_n)\).  For the dyadic partition, the remaining weighted
logarithmic term can be controlled at the ensemble scale using a mixed
logarithmic occupation bound.  On \(\{x_0|B_n(x_0)>0\}\), the two crossing
directions alternate; therefore,
\(\theta_{1,n}=1/2+O(B_n^{-1})\) and
\(\theta_{2,n}=1/2+O(B_n^{-1})\).  More generally,
\(-\sum_i\theta_{i,n}\log\theta_{i,n}\le \log2\), and hence
\begin{equation}
\left|
 p_n\sum_{i=1}^{2}\theta_{i,n}
 \log\frac{\rho_i^{(n)}}{\theta_{i,n}}
\right|
\le
p_n\left(
\log2+
\sum_{i=1}^{2}\theta_{i,n}|\log\rho_i^{(n)}|
\right),
\label{eq:weighted_log_bound_dyadic}
\end{equation}
where terms with a zero prefactor are omitted, and the entire expression is set to zero when \(p_n=0\).  We assume the following mixed logarithmic occupation control:
\begin{equation}
\left\langle
p_n\left(
1+
\sum_{i=1}^{2}\theta_{i,n}|\log\rho_i^{(n)}|
\right)
\right\rangle_{\rm ens}
=
O\!\left(\left\langle p_n\right\rangle_{\rm ens}\right).
\label{eq:mixed_log_occupation_assumption}
\end{equation}
For the dyadic partition, \(\rho_1^{(n)}\) and \(\rho_2^{(n)}\) are occupation
fractions of the two laminar components and not the rare-transition frequencies.
Therefore, the assumption in Eq.~\eqref{eq:mixed_log_occupation_assumption}
requires that the logarithms of these occupation fractions do not
produce an additional factor of order \(L_n=\log(n/a_n)\) after multiplication
by \(p_n\) and ensemble averaging. Under
Eq.~\eqref{eq:mixed_log_occupation_assumption}, the weighted logarithmic term
has an ensemble average of order \(\langle p_n\rangle_{\rm ens}\).  Together with
the bound on the diagonal (self-transition) terms, this implies that the
non-leading contributions are absorbed into partition-dependent coefficients.
Consequently, the ensemble scale of the empirical entropy is determined as
\begin{equation*}
    \left\langle H_{P_2}^{(1,n)}\right\rangle_{\rm ens}
    =
    O\!\left(\left\langle p_n\log\frac{1}{p_n}\right\rangle_{\rm ens}\right)
    +
    O\!\left(\left\langle p_n\right\rangle_{\rm ens}\right),
\end{equation*}
up to partition-dependent constants.

We now estimate these two averages.  Because \(p_n=B_n/n\), where \(B_n\) is
the number of cross transitions, estimating the scale of \(B_n\) is sufficient.
We define the crossing set
\begin{equation*}
    C_2=\{x:\sigma(Tx)\ne\sigma(x)\}
\end{equation*}
and let
\begin{equation*}
    g_2(x)=\mathbf{1}_{C_2}(x).
\end{equation*}
Therefore,
\begin{equation*}
    B_n(x_0)=\sum_{k=0}^{n-1}g_2(T^kx_0)
    =
    \sum_{k=0}^{n-1}\mathbf{1}_{C_2}(T^kx_0).
\end{equation*}
Thus, \(B_n\) is the exact number of visits to the crossing set \(C_2\).  For
the maps and partitions considered herein, \(C_2\) is separated from the
indifferent fixed points and has a finite invariant measure.  Hence
\(g_2\in L^1(\mu)\).  Under the corresponding Darling--Kac--Aaronson return-sequence scaling for 
this finite-measure indicator, \(B_n\) is of return-sequence order \(a_n\) in distribution.  
In the following ensemble estimates, we use the corresponding
mean scaling
\begin{equation}
    \left\langle B_n\right\rangle_{\rm ens}\asymp a_n.
    \label{eq:Bn_mean_order}
\end{equation}
Henceforth, \(f_n\asymp g_n\) implies that the constants
\(C_-,C_+>0\) and \(n_0\) exist, such that
\begin{equation*}
    C_-g_n\le f_n\le C_+g_n
    \qquad (n\ge n_0).
\end{equation*}
Thus,
\begin{equation}
    \left\langle p_n\right\rangle_{\rm ens}
    =
    \frac{\left\langle B_n\right\rangle_{\rm ens}}{n}
    \asymp
    \frac{a_n}{n}.
    \label{eq:pn_mean_order}
\end{equation}

We now estimate the logarithmic average.  The normalized crossing
count is defined by
\begin{equation*}
    Z_n=\frac{B_n}{a_n}.
\end{equation*}
Thus,
\begin{equation*}
    p_n=\frac{B_n}{n}=\frac{a_n}{n}Z_n.
\end{equation*}
Substituting this into \(-p_n\log p_n\), we obtain
\begin{align*}
    -p_n\log p_n
    &=
    -\frac{a_n}{n}Z_n
    \log\left(\frac{a_n}{n}Z_n\right)
    \\
    &=
    -\frac{a_n}{n}Z_n
    \left(\log\frac{a_n}{n}+\log Z_n\right)
    \\
    &=
    \frac{a_n}{n}Z_n
    \left(\log\frac{n}{a_n}-\log Z_n\right).
\end{align*}
With
\begin{equation*}
    L_n=\log\frac{n}{a_n},
\end{equation*}
this becomes
\begin{equation*}
    -p_n\log p_n
    =
    \frac{a_n}{n}\left(Z_nL_n-Z_n\log Z_n\right).
\end{equation*}
Taking the ensemble average yields
\begin{equation*}
    \left\langle -p_n\log p_n\right\rangle_{\rm ens}
    =
    \frac{a_n}{n}
    \left(
        L_n\left\langle Z_n\right\rangle_{\rm ens}
        -
        \left\langle Z_n\log Z_n\right\rangle_{\rm ens}
    \right).
\end{equation*}
This moment control is natural for the dyadic partition because \(Z_n\) is
proportional to the scaled correlation defect.  From
Eq.~\eqref{eq:pn_cross_frequency},
\begin{equation}
    Z_n
    =
    \frac{B_n}{a_n}
    =
    \frac{1}{2}\,\frac{n}{a_n}\left(1-C(1,n)\right).
    \label{eq:Zn_correlation_defect}
\end{equation}
Similarly, if
\(\eta_1(x)=1-\sigma(x)\sigma(Tx)\), then
\(\sum_{k=0}^{n-1}\eta_1(T^kx_0)=2B_n(x_0)\).  For the modified Bernoulli map,
according to Ref. \onlinecite{akimoto2007new}, it holds that
\begin{equation}
    \frac{n}{a_n}\left(1-C(1,n)\right)
    \overset{d}{\longrightarrow}
    \left(\int \eta_1\,d\mu\right)Y_\alpha,
    \label{eq:akimoto_correlation_limit}
\end{equation}
where \(Y_\alpha\) denotes the normalized Mittag--Leffler limit variable.
Thus,
\begin{equation*}
    Z_n
    \overset{d}{\longrightarrow}
    \frac{1}{2}\left(\int \eta_1\,d\mu\right)Y_\alpha.
\end{equation*}
This distributional limit motivates the following moment assumption: For the
Boole transformation, we use it as the analogous Darling--Kac-Aaronson type moment
control.  We assume that
\(\langle Z_n\rangle_{\rm ens}=O(1)\) with nonzero limiting order and that the
logarithmic moment is uniformly controlled,
\(\langle |Z_n\log Z_n|\rangle_{\rm ens}=O(1)\), with the convention
\(Z_n\log Z_n=0\) when \(Z_n=0\).  Subsequently,
\begin{equation}
    \left\langle -p_n\log p_n\right\rangle_{\rm ens}
    \asymp
    \frac{a_n}{n}L_n.
    \label{eq:pn_log_mean_order}
\end{equation}
Combining Eqs.~\eqref{eq:pn_mean_order} and \eqref{eq:pn_log_mean_order}, we
obtain the two-term finite-time estimate
\begin{equation}
    \left\langle H_P^{(1,n)}\right\rangle_{\rm ens}
    \simeq
    c_{1,P}\frac{a_n}{n}L_n
    +
    c_{2,P}\frac{a_n}{n}.
    \label{eq:HP_two_term_mean}
\end{equation}
Here \(c_{1,P}\) and \(c_{2,P}\) are effective constants that depend on the
fixed partition \(P\).  For the fixed cylinder partitions used below, we use
the same two-term form with partition-dependent coefficients.

Similarly, for the information sum \(I_P(n)=nH_P^{(1,n)}\),
\begin{equation}
    \left\langle I_P(n)\right\rangle_{\rm ens}
    \simeq
    c_{1,P}a_nL_n+c_{2,P}a_n.
    \label{eq:IP_two_term_mean}
\end{equation}
Therefore, the leading-order decay of the per-step quantity is 
\begin{equation}
    \left\langle H_P^{(1,n)}\right\rangle_{\rm ens}
    =
    O\!\left(\frac{a_n}{n}\log\frac{n}{a_n}\right),
    \label{eq:HP_leading_order}
\end{equation}
with an \(O(a_n/n)\) correction, which is important in finite-time simulations,
particularly near the transition to the infinite-measure regime, where \(L_n\)
grows slowly.

For the modified Bernoulli map with parameter \(B_{\rm MB}>2\),
\[
    a_n\asymp n^\alpha,
    \qquad
    \alpha=\frac{1}{B_{\rm MB}-1}.
\]
For the Boole transformation, \(a_n\asymp n^{1/2}\).  In the numerical
comparisons in the following, we use \(a_n=n^\alpha\); the unknown multiplicative constant
in the true return sequence is absorbed into the fitted coefficients
\(c_{1,P}\) and \(c_{2,P}\).  We use fixed cylinder partitions with
\(|P|=2,4,8\).

\begin{figure*}
    \centering
    \includegraphics[width=0.90\textwidth]{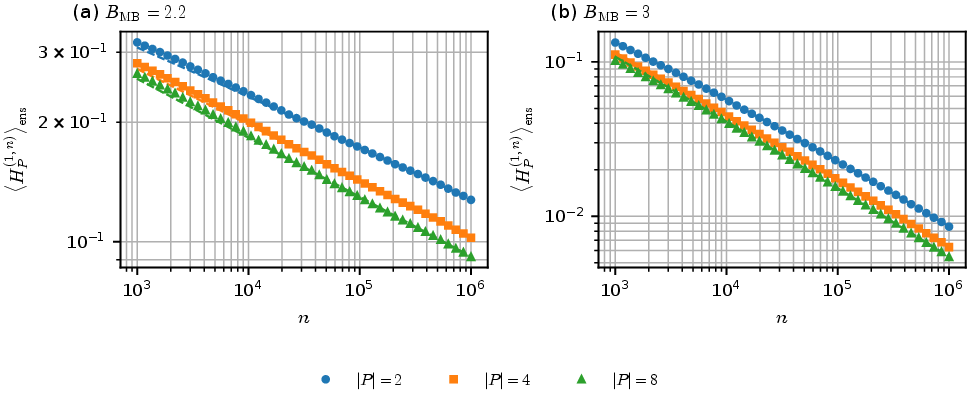}
    \caption{Time evolution of the ensemble mean
    \(\langle H_P^{(1,n)}\rangle_{\rm ens}\) for the modified Bernoulli map.
    The symbols show numerical results for fixed cylinder partitions with
    \(|P|=2,4,8\), and the dashed curves show fits based on the two-term form in
    Eq.~\eqref{eq:HP_two_term_mean}.  The parameter \(B_{\rm MB}=2.2\) is
    closer to the finite-measure transition and shows stronger finite-time
    correction, whereas \(B_{\rm MB}=3\) displays the leading
    sparse-transition scale more clearly.  Initial conditions were sampled independently from the Lebesgue distribution
on \((0,1)\); the ensemble size was \(M=5000\).  The plotted horizon \(n\)
denotes the number of empirical cylinder-state transitions, with
\(n=10^6\).  For the largest partition shown, \(|P|=8\) (depth
\(m=3\)), constructing the cylinder labels requires the underlying scalar
orbit up to \(x_{n+2}\).}
    \label{fig:section4_MB_H_vs_n}
\end{figure*}

\begin{figure*}
    \centering
    \includegraphics[width=0.90\textwidth]{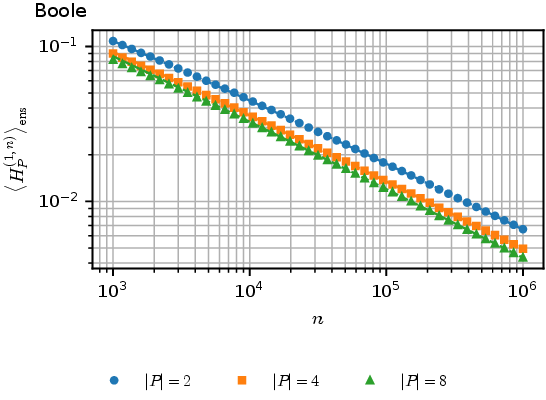}
    \caption{Time evolution of the ensemble mean
    \(\langle H_P^{(1,n)}\rangle_{\rm ens}\) for the Boole transformation.
    The symbols show numerical results for fixed cylinder partitions with
    \(|P|=2,4,8\), and the dashed curves show fits based on
    Eq.~\eqref{eq:HP_two_term_mean} with \(\alpha=1/2\).  Initial conditions were sampled independently from the Lebesgue distribution
on \((0,1)\); the ensemble size was \(M=5000\).  The plotted horizon \(n\)
denotes the number of empirical cylinder-state transitions, with
\(n=10^6\).  For the largest partition shown, \(|P|=8\) (depth
\(m=3\)), constructing the cylinder labels requires the underlying scalar
orbit up to \(x_{n+2}\).}
    \label{fig:section4_Boole_H_vs_n}
\end{figure*}

Figure~\ref{fig:section4_MB_H_vs_n} shows
\(\langle H_P^{(1,n)}\rangle_{\rm ens}\) for the modified Bernoulli map.  The
case \(B_{\rm MB}=2.2\) illustrates the importance of the
\(c_{2,P}a_n/n\) term at accessible times, whereas for \(B_{\rm MB}=3\), the
leading logarithmic scale is already more visible.  Figure~\ref{fig:section4_Boole_H_vs_n}
shows the corresponding results for the Boole transformation.  In all scenarios,
the curves are consistent with the decay predicted by
Eq.~\eqref{eq:HP_two_term_mean}.

\section{Self-normalized fluctuations of the one-step conditional entropy}
\label{sec:self_normalized_fluctuations}

The preceding sections show two facts that may at first appear contradictory.  On the one hand,
\[
    H_P^{(1,n)}(x_0)\to0
\]
for fixed partitions in the infinite-measure regime.  On the other hand, the finite-time information sum \(I_P(n,x_0)=nH_P^{(1,n)}(x_0)=\sum_{k=0}^{n-1}\phi_P^{(n)}(x_k)\) is controlled by rare intermittent transitions and has a nontrivial sublinear growth scale.  This suggests that the raw value of \(H_P^{(1,n)}\) is not the correct object for detecting weak chaos; however, its trajectory-to-trajectory fluctuations may still contain meaningful information.

This motivation originates from the Darling--Kac--Aaronson distributional picture.  In the Darling--Kac--Aaronson setting, for a positive observable
\(f\in L^1(\mu)\) satisfying the assumptions of the theorem, one has
\begin{equation}
    \frac{1}{a_n}
    \sum_{k=0}^{n-1} f(T^kx)
    \overset{d}{\longrightarrow}
    \left(\int f\,d\mu\right)Y_\alpha,
    \label{Eq: DKA theorem}
\end{equation}
where \(Y_\alpha\) is a normalized Mittag--Leffler random variable of index \(\alpha\in(0,1)\) \cite{aaronson1997introduction}.  Strictly, Eq.~\eqref{Eq: DKA theorem} does not apply directly to \(I_P(n)\) because the summand \(\phi_P^{(n)}\) depends on \(n\) through the empirical transition matrix.  However, the sparse-transition analysis presented in Sec.~IV shows that \(I_P(n)\) is largely controlled by the same intermittent return activity that governs
Birkhoff sums of fixed \(L^1(\mu)\) observables in the
Darling--Kac--Aaronson framework.  Therefore, it is natural to ask whether the normalized fluctuations exhibit a Mittag--Leffler-type shape.

The scaling described in Sec.~IV also shows that the pure return-sequence normalization \(I_P(n)/a_n\) is not an appropriate absolute scale for the empirical Markov information sum: its ensemble mean increases as \(c_{1,P}L_n+c_{2,P}\).  Therefore, if we seek to introduce an auxiliary finite-time normalization for \(I_P(n)\), the natural leading scale is not \(a_n\) alone but
\[
    a_nL_n,
    \qquad
    L_n=\log\frac{n}{a_n}.
\]
Thus, we define the leading sparse-transition-scale-normalized information sum
\begin{equation}
    \widehat H_P^{(1,n)}(x_0)
    \equiv
    \frac{1}{a_nL_n}
    \sum_{k=0}^{n-1}\phi_P^{(n)}(x_k)
    =
    \frac{I_P(n,x_0)}{a_nL_n}
    =
    \frac{n}{a_nL_n}H_P^{(1,n)}(x_0).
    \label{eq:Hhat_sparse_scale}
\end{equation}
Through Eq.~\eqref{eq:IP_two_term_mean}, this normalization yields
\begin{equation}
    \left\langle \widehat H_P^{(1,n)}\right\rangle_{\rm ens}
    \simeq
    c_{1,P}+\frac{c_{2,P}}{L_n}.
    \label{eq:Hhat_mean}
\end{equation}
Thus, \(\widehat H_P^{(1,n)}\) has an \(O(1)\) ensemble mean at the level of the leading sparse-transition scale up to the finite-time \(O(1/L_n)\) correction.  However, we stress that \(\widehat H_P^{(1,n)}\) is still not a new entropy rate or a direct DKA observable because \(\phi_P^{(n)}\) depends on the empirical transition matrix at time \(n\).

The useful point in this section is that this deterministic normalization cancels out after division by the ensemble mean:
\begin{equation}
    \frac{\widehat H_P^{(1,n)}(x_0)}
    {\left\langle \widehat H_P^{(1,n)}\right\rangle_{\rm ens}}
    =
    \frac{H_P^{(1,n)}(x_0)}
    {\left\langle H_P^{(1,n)}\right\rangle_{\rm ens}}
    =
    \frac{I_P(n,x_0)}
    {\left\langle I_P(n)\right\rangle_{\rm ens}}.
    \label{eq:self_normalization_identity}
\end{equation}
Thus, the object of interest is the self-normalized variable
\begin{equation}
    Z_P^{(n)}(x_0)
    \equiv
    \frac{H_P^{(1,n)}(x_0)}
    {\left\langle H_P^{(1,n)}\right\rangle_{\rm ens}}
    =
    \frac{I_P(n,x_0)}
    {\left\langle I_P(n)\right\rangle_{\rm ens}}.
    \label{eq:ZP_definition}
\end{equation}
This definition does not change the per-step normalization of \(H_P^{(1,n)}\).  Instead, it removes the deterministic decay of the ensemble mean and isolates the remaining finite-time fluctuations.

The density of the normalized Mittag--Leffler random variable \(Y_\alpha\) is denoted by \(M_\alpha(y)\) and is normalized such that \(\langle Y_\alpha\rangle=1\).  In the following numerical comparisons, we compare the empirical distribution of \(Z_P^{(n)}\) with \(M_\alpha\), where \(\alpha=1/(B_{\rm MB}-1)\) for the modified Bernoulli map and \(\alpha=1/2\) for the Boole transformation.

For the distributional comparison, we use moderately refined fixed cylinder partitions, \(|P|=8,16,32\).  This choice is made for finite-time numerical stability and not as a conceptual change of the problem.  For the coarsest dyadic partition \(|P|=2\), the information sum is dominated by a single rare crossing count.  By using fixed cylinder partitions with a moderately larger number of atoms, the empirical information sum samples a richer set of transition types, and the self-normalized distribution becomes more stable.  More importantly, the partition size is fixed as \(n\) varies.

\begin{figure*}
    \centering
    \includegraphics[width=0.90\textwidth]{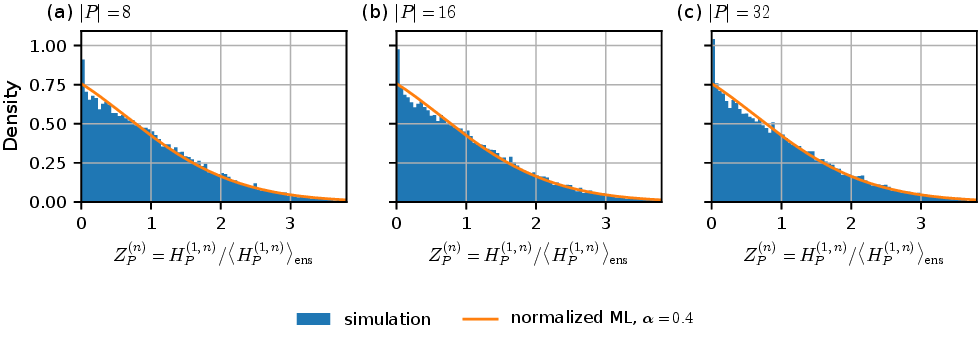}
    \caption{Self-normalized distribution \(Z_P^{(n)}=H_P^{(1,n)}/\langle H_P^{(1,n)}\rangle_{\rm ens}\) for the modified Bernoulli map with \(B_{\rm MB}=3.5\), corresponding to \(\alpha=0.4\).  Fixed cylinder partitions \(|P|=8,16,32\) are used.  The solid curve is the normalized Mittag--Leffler density with \(\alpha=0.4\).  Initial conditions were sampled independently from the Lebesgue distribution
on \((0,1)\); the ensemble size was \(M=30000\).  The horizon
\(n=10^5\) denotes the number of empirical cylinder-state transitions.
For the largest partition shown, \(|P|=32\) (depth \(m=5\)), constructing
the cylinder labels requires the underlying scalar orbit up to \(x_{n+4}\).}
    \label{fig:section5_MB_B35}
\end{figure*}

\begin{figure*}
    \centering
    \includegraphics[width=0.90\textwidth]{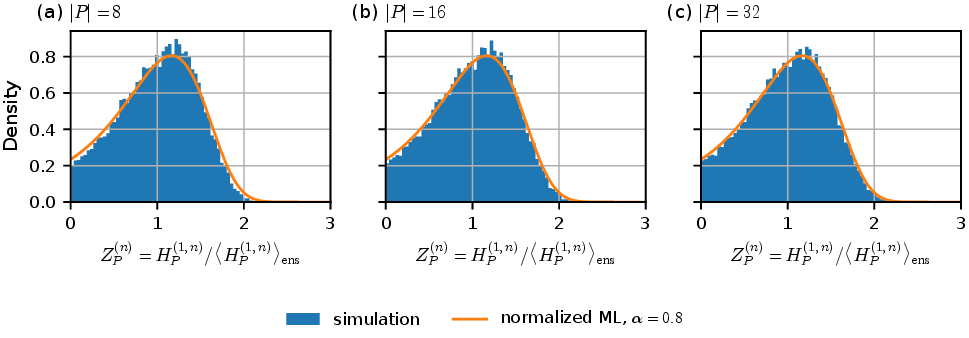}
    \caption{Self-normalized distribution \(Z_P^{(n)}=H_P^{(1,n)}/\langle H_P^{(1,n)}\rangle_{\rm ens}\) for the modified Bernoulli map with \(B_{\rm MB}=2.25\), corresponding to \(\alpha=0.8\).  Fixed cylinder partitions \(|P|=8,16,32\) are used. The solid curve is the normalized Mittag--Leffler density with \(\alpha=0.8\). Initial conditions were sampled independently from the Lebesgue distribution
on \((0,1)\); the ensemble size was \(M=30000\).  The horizon
\(n=10^5\) denotes the number of empirical cylinder-state transitions.
For the largest partition shown, \(|P|=32\) (depth \(m=5\)), constructing
the cylinder labels requires the underlying scalar orbit up to \(x_{n+4}\).}
    \label{fig:section5_MB_B225}
\end{figure*}

\begin{figure*}
    \centering
    \includegraphics[width=0.90\textwidth]{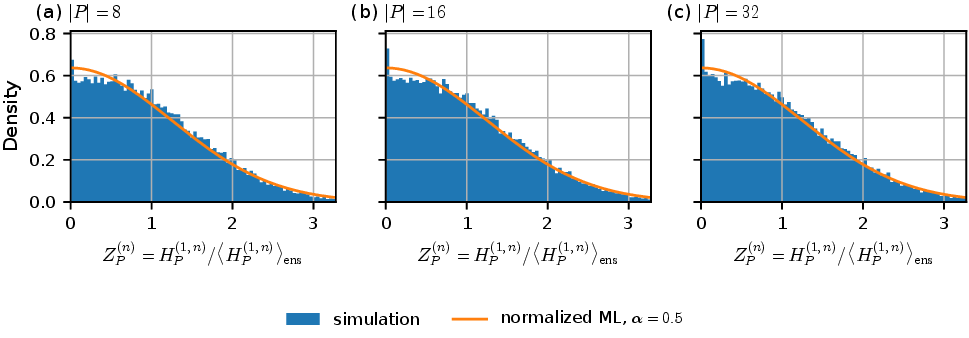}
    \caption{Self-normalized distribution \(Z_P^{(n)}=H_P^{(1,n)}/\langle H_P^{(1,n)}\rangle_{\rm ens}\) for the Boole transformation, corresponding to $\alpha=1/2$. Fixed cylinder partitions \(|P|=8,16,32\) are used. The solid curve is the normalized Mittag--Leffler density with $\alpha=1/2$.     Initial conditions were sampled independently from the Lebesgue distribution
on \((0,1)\); the ensemble size was \(M=30000\).  The horizon
\(n=10^5\) denotes the number of empirical cylinder-state transitions.
For the largest partition shown, \(|P|=32\) (depth \(m=5\)), constructing
the cylinder labels requires the underlying scalar orbit up to \(x_{n+4}\).}
    \label{fig:section5_Boole}
\end{figure*}

Figures~\ref{fig:section5_MB_B35}, \ref{fig:section5_MB_B225}, and \ref{fig:section5_Boole} present the distributions of \(Z_P^{(n)}\).  For the modified Bernoulli map, we use \(B_{\rm MB}=3.5\) (\(\alpha=0.4\)) and \(B_{\rm MB}=2.25\) (\(\alpha=0.8\)).  For the Boole transformation, \(\alpha=1/2\).  In all scenarios, the self-normalized histograms are consistent with the corresponding
normalized Mittag--Leffler density, with residual finite-time deviations depending on the map and partition resolution.

The comparisons shown in Figs.~\ref{fig:section5_MB_B35}--\ref{fig:section5_Boole} use fixed cylinder partitions.
This choice is convenient for controlling the symbolic
dynamics and comparing different partition resolutions.
For a complementary robustness check, Appendix~\ref{app:equal-bin}
uses a map-blind equal-bin construction.  Given only the
scalar trajectory data, a fixed observed range is divided into
\(K\) equal bins, and the same empirical one-step transition
matrix is used to compute \(H_P^{(1,n)}\).  The resulting
self-normalized distributions are again consistent with the
corresponding normalized Mittag--Leffler densities.  This
suggests that the Mittag--Leffler-type fluctuation observed
in \(Z_P^{(n)}\) is not an artifact of the cylinder partitions
used in the main figures but reflects the intermittent return
activity observed through a fixed finite symbolic observation.

\section{Relation to generalized Lyapunov sums:
sparse-transition information sums measured relative to \(G_n\)}
\label{sec:relation_gle_sparse}

In Sec.~V, we study the self-normalized variable
\begin{equation}
    Z_P^{(n)}
    =
    \frac{H_P^{(1,n)}}{\langle H_P^{(1,n)}\rangle_{\rm ens}} .
    \label{eq:sec6_ZP_def}
\end{equation}
In this normalization, any deterministic prefactor is canceled.  In this section,
we address a different question: how is the finite-time information sum
\begin{equation}
    I_P(n,x_0)
    \equiv
    nH_P^{(1,n)}(x_0)
    =
    \sum_{k=0}^{n-1}\phi_P^{(n)}(x_k)
    \label{eq:sec6_IP_def}
\end{equation}
related to existing weak-chaos indicators?  The aim is not to propose
another generalized Pesin identity.  Rather, we clarify why the empirical one-step Markov 
information sum grows on a scale different from that of generalized Lyapunov sums and 
Lempel--Ziv complexity, and then show that it is nevertheless controlled by the same 
finite-time intermittent activity as measured by the generalized Lyapunov sum.

We define the finite-time generalized Lyapunov sum as
\begin{equation}
    G_n(x_0)
    \equiv
    \sum_{k=0}^{n-1}\log |T'(x_k)| .
    \label{eq:Gn_def}
\end{equation}
The corresponding finite-time generalized Lyapunov exponent is obtained by
normalizing this sum by using the return sequence
\begin{equation}
    \lambda_\alpha^{(n)}(x_0)
    \equiv
    \frac{G_n(x_0)}{a_n} .
    \label{eq:lambda_alpha_n_def}
\end{equation}
For intermittent maps with infinite invariant densities, such
return-sequence-normalized Lyapunov sums are known to display
Mittag--Leffler fluctuations and to be related to Krengel entropy through
Pesin-type identities \cite{akimoto2007new,korabel2009pesin,korabel2010separation}.  In
contrast, the observable entering \(I_P(n)\) is not a fixed \(L^1(\mu)\)
observable.  The one-step conditional self-information \(\phi_P^{(n)}\) depends on the
time horizon \(n\) through the empirical transition matrix. 

The scaling analysis described in Sec.~IV yields, in the infinite-measure regime,
\begin{equation}
    \langle I_P(n)\rangle_{\rm ens}
    =
    \langle nH_P^{(1,n)}\rangle_{\rm ens}
    =
    O\!\left(a_n\log\frac{n}{a_n}\right)+O(a_n).
    \label{eq:IP_alog_plus_a_sec6}
\end{equation}
In contrast, \(G_n\) is a Birkhoff sum of a fixed integrable observable and
belongs to the return-sequence scale.  The additional logarithmic factor in
Eq.~\eqref{eq:IP_alog_plus_a_sec6} has a direct information-theoretic
interpretation.  In the infinite-measure regime, the transitions between long
laminar phases are rare, with an empirical frequency of the order \(a_n/n\).  A
fixed one-step empirical Markov model assigns a rare-transition self-information factor of the order
\(\log(n/a_n)\) to such rare transitions.  Thus,
\begin{widetext}
\begin{equation}
    \text{empirical Markov information sum}
    \sim
    \text{number of rare transitions}
    \times
    \text{self-information per rare transition}
    \sim
    a_n\log\frac{n}{a_n}.
    \label{eq:total_markov_self_info_scaling}
\end{equation}
\end{widetext}
Here the phrase ``empirical Markov information sum'' refers to
\[
    I_P(n,x_0)
    =
    \sum_{i,j}N_{ij}^{(n)}(x_0)
    \bigl[-\log t_{ij}^{(n)}(x_0)\bigr],
\]
that is, the self-information sum assigned by the empirical one-step Markov
transition model.  Lempel--Ziv complexity is related to
the return-sequence class \cite{korabel2010separation}.

We now explain why \(I_P(n)\) should be compared with \(G_n\).
Let \(P\) be a fixed cylinder partition and let \(s_k^{(P)}\) denote
the corresponding symbols at time \(k\).  Similarly, we let \(s^{(P)}(x)\) denote
the symbol of the partition atom containing \(x\).  We define the indicator of a
non-self-transition by
\begin{equation}
    g_P(x)
    \equiv
    {\bf 1}_{\{s^{(P)}(Tx)\ne s^{(P)}(x)\}},
    \label{eq:gP_def}
\end{equation}
and let
\begin{equation}
    B_{P,n}(x_0)
    \equiv
    \sum_{k=0}^{n-1}g_P(T^kx_0)
    =
    \#\{k|s_{k+1}^{(P)}\ne s_k^{(P)},0\le k\le n-1\} .
    \label{eq:BPn_def}
\end{equation}
Thus, \(B_{P,n}\) is the total number of non-self-symbolic transitions up to
time \(n\).  For the dyadic partition
\begin{equation}
    P_2=\{[0,1/2),[1/2,1]\},
    \label{eq:dyadic_partition_sec6}
\end{equation}
\(B_{P_2,n}\) coincides with the cross-transition count used in Sec.~IV.

For the fixed dyadic and cylinder partitions considered here, \(g_P\)
vanishes in sufficiently small neighborhoods of the indifferent fixed points
and is integrable with respect to the infinite invariant measure.  The
Lyapunov observable
\begin{equation}
    f(x)=\log |T'(x)|
    \label{eq:lyap_observable_f}
\end{equation}
is also integrable with respect to the same infinite measure because it
vanishes at the indifferent fixed points, where the invariant density diverges.
Hence, Hopf's ratio ergodic theorem \cite{aaronson1997introduction} suggests
that, for typical initial conditions in which the denominator diverges,
\begin{equation}
    \frac{B_{P,n}(x_0)}{G_n(x_0)}
    =
    \frac{\sum_{k=0}^{n-1}g_P(T^kx_0)}
    {\sum_{k=0}^{n-1}f(T^kx_0)}
    \longrightarrow
    \xi_P
    \equiv
    \frac{\int g_P\,d\mu}{\int f\,d\mu} .
    \label{eq:Hopf_ratio_BP_G}
\end{equation}
Here, this theorem is used only as a guide; it applies directly to fixed
integrable observables, whereas \(I_P(n)\) also contains empirical
transition probabilities depending on \(n\).  Equation~\eqref{eq:Hopf_ratio_BP_G}
gives the theoretical basis for expecting the transition count \(B_{P,n}\)
and the generalized Lyapunov sum \(G_n\) to be governed by the same
finite-time intermittent activity.

Next, we connect this transition count to the empirical Markov information sum.
By definition,
\begin{equation}
    I_P(n,x_0)
    =
    \sum_{i,j}N_{ij}^{(n)}(x_0)
    \log\frac{N_i^{(n)}(x_0)}{N_{ij}^{(n)}(x_0)} .
    \label{eq:IP_counts_sec6}
\end{equation}
Let \(\mathcal R_P\) denote the set of non-self-transition types,
\[
    \mathcal R_P=\{(i,j):i\ne j\},
\]
restricted to the admissible transitions of the fixed partition \(P\).  In
the intermittent regime, these non-self-transitions are rare.  By construction,
\begin{equation}
    B_{P,n}
    =
    \sum_{(i,j)\in\mathcal R_P}N_{ij}^{(n)} .
    \label{eq:BPn_sum_nonself}
\end{equation}
For the dyadic partition, \(\mathcal R_{P_2}=\{(1,2),(2,1)\}\), and this
definition reduces to the cross-transition count.

For \(B_{P,n}>0\), we define
\begin{equation}
    \theta_{ij,n}
    \equiv
    \frac{N_{ij}^{(n)}}{B_{P,n}},
    \qquad
    \rho_i^{(n)}
    \equiv
    \frac{N_i^{(n)}}{n},
    \qquad
    (i,j)\in\mathcal R_P .
    \label{eq:theta_rho_sec6_def}
\end{equation}
Here, \(\theta_{ij,n}\) is the fraction of all non-self-transitions that have
type \(X_i\to X_j\), and \(\rho_i^{(n)}\) is the empirical occupation frequency of
the source state \(i\).  Because \(B_{P,n}\) is the sum of all non-self-
transition counts, we obtain
\begin{equation}
    \sum_{(i,j)\in\mathcal R_P}\theta_{ij,n}=1 .
    \label{eq:theta_sum_one_sec6}
\end{equation}
If \(B_{P,n}=0\), no non-self-transition is observed.  If only
self-transitions occur along the observed symbolic trajectory, 
\(I_P(n,x_0)=0\).  The term \(B_{P,n}\log(n/B_{P,n})\) is then defined by
its continuous extension at \(B_{P,n}=0\), that is, zero.

With this notation, the non-self-transition part of Eq.~\eqref{eq:IP_counts_sec6}
is decomposed as
\begin{align}
    \sum_{(i,j)\in\mathcal R_P}N_{ij}^{(n)}
    \log\frac{N_i^{(n)}}{N_{ij}^{(n)}}
    &=
    B_{P,n}\log\frac{n}{B_{P,n}}
    \nonumber\\
    &\quad+
    B_{P,n}
    \sum_{(i,j)\in\mathcal R_P}
    \theta_{ij,n}
    \log\frac{\rho_i^{(n)}}{\theta_{ij,n}} .
    \label{eq:rare_part_decomposition}
\end{align}
The first term,
\[
    B_{P,n}\log\frac{n}{B_{P,n}},
\]
is a coarse-grained sparse-transition contribution: It treats all non-self
transitions as one rare class of empirical frequency \(B_{P,n}/n\) and
assigns the coarse-grained self-information
\(\log(n/B_{P,n})\) to each such transition.  The
second term contains the correction due to the distribution of non-self
transitions among individual transition types and occupation
frequencies of the corresponding source states.  For the dyadic partition
this correction is relatively simple, because only the two crossing types
\(X_1\to X_2\) and \(X_2\to X_1\) contribute.  For deeper cylinder partitions, several
non-self-transition types contribute, and the source-state occupation
frequencies may produce significant finite-time corrections.  These effects are
absorbed into the partition-dependent effective coefficients described in the following.

It remains to estimate the diagonal, or self-transition, contribution in
Eq.~\eqref{eq:IP_counts_sec6}.  For each state \(i\), let
\[
    Q_i^{(n)}
    \equiv
    \sum_{j\ne i}N_{ij}^{(n)}
\]
be the number of non-self-transitions leaving state \(i\).  As
\[
    N_i^{(n)}
    =
    N_{ii}^{(n)}+Q_i^{(n)},
\]
the diagonal contribution is
\begin{align}
    D_P(n,x_0)
    &\equiv
    \sum_i
    N_{ii}^{(n)}
    \log\frac{N_i^{(n)}}{N_{ii}^{(n)}}     \nonumber\\
    &=
    \sum_i
    \bigl(N_i^{(n)}-Q_i^{(n)}\bigr)
    \log
    \frac{N_i^{(n)}}{N_i^{(n)}-Q_i^{(n)}} ,
    \label{eq:diagonal_DP_def}
\end{align}
where the terms with zero prefactor are understood through continuity.  For
\(0\le Q\le N\), the elementary inequality
\begin{equation}
    0\le
    (N-Q)\log\frac{N}{N-Q}
    \le Q
    \label{eq:diagonal_elementary_ineq}
\end{equation}
follows from \(\log x\le x-1\) with \(x=N/(N-Q)\).  Therefore,
\begin{equation}
    0\le
    D_P(n,x_0)
    \le
    \sum_i Q_i^{(n)}
    =
    \sum_{i\ne j}N_{ij}^{(n)}
    =
    B_{P,n}(x_0).
    \label{eq:diagonal_DP_bound}
\end{equation}
Thus, the diagonal contribution is bounded by the total number of non-self-transitions.

Combining Eqs.~\eqref{eq:rare_part_decomposition} and
\eqref{eq:diagonal_DP_def}, we obtain, for \(B_{P,n}>0\), the finite-time
representation
\begin{equation}
    I_P(n,x_0)
    =
    B_{P,n}(x_0)\log\frac{n}{B_{P,n}(x_0)}
    +
    \Gamma_P(n,x_0)B_{P,n}(x_0),
    \label{eq:IP_BlogB_expansion}
\end{equation}
where
\begin{equation}
    \Gamma_P(n,x_0)
    =
    \sum_{(i,j)\in\mathcal R_P}
    \theta_{ij,n}
    \log\frac{\rho_i^{(n)}}{\theta_{ij,n}}
    +
    \frac{D_P(n,x_0)}{B_{P,n}(x_0)} .
    \label{eq:GammaP_def}
\end{equation}
This representation separates the contribution obtained by treating all
non-self-transitions as one rare class from the correction that depends on the
transition-type distribution, source-state occupation frequencies, and 
diagonal self-transition part.  For a dyadic partition, this lower-order statement can be quantified at the
ensemble-scaling level.  As the two crossing directions alternate, and
\(0\le D_{P_2}/B_{P_2,n}\le1\), we have
\[
|\Gamma_{P_2}(n,x_0)|
\le
1+\log2+
\sum_{i=1}^{2}\theta_{i,n}|\log\rho_i^{(n)}(x_0)|.
\]
Using the same assumptions as those in
Eq.~\eqref{eq:mixed_log_occupation_assumption}, together with
\(\langle B_{P_2,n}\rangle_{\rm ens}\asymp a_n\), we obtained
\[
\left\langle
|B_{P_2,n}\Gamma_{P_2}|
\right\rangle_{\rm ens}
=O(a_n).
\]
Similarly,
\[
\left\langle
\left|B_{P_2,n}\log\frac{B_{P_2,n}}{a_n}\right|
\right\rangle_{\rm ens}
=
a_n\left\langle |Z_n\log Z_n|\right\rangle_{\rm ens}
=O(a_n).
\]
Thus, for the dyadic partition, the bracketed correction in
Eq.~\eqref{eq:IP_BLn_Gamma_form} contributes at the order \(a_n\), whereas the
leading sparse-transition term has the order \(a_nL_n\).  However, for deeper cylinder partitions,
some source-state occupation frequencies may be small, and the logarithms of
\(\rho_i^{(n)}\) in \(\Gamma_P\) can produce corrections comparable to
\(L_n=\log(n/a_n)\) at accessible times.  Therefore,
Eq.~\eqref{eq:IP_BlogB_expansion} should not be interpreted as implying that
\(I_P/[B_{P,n}\log(n/B_{P,n})]\) is close to unity for all fixed partitions.

Using
\begin{equation}
    \log\frac{n}{B_{P,n}}
    =
    \log\frac{n}{a_n}
    -
    \log\frac{B_{P,n}}{a_n}
    =
    L_n-
    \log\frac{B_{P,n}}{a_n},
    \qquad
    L_n\equiv\log\frac{n}{a_n},
    \label{eq:log_n_over_B_decomp}
\end{equation}
Eq.~\eqref{eq:IP_BlogB_expansion} becomes
\begin{equation}
    I_P(n,x_0)
    =
    B_{P,n}(x_0)L_n
    +
    B_{P,n}(x_0)
    \left[
        \Gamma_P(n,x_0)
        -
        \log\frac{B_{P,n}(x_0)}{a_n}
    \right].
    \label{eq:IP_BLn_Gamma_form}
\end{equation}
The first term explicitly separates the deterministic logarithmic factor
\(L_n\).  The bracketed term contains the finite-time corrections from
the distribution of non-self-transition types, the source-state occupation
frequencies, the diagonal self-transition contribution, and the fluctuation of
\(B_{P,n}/a_n\).  For the dyadic partition, this bracketed term is expected to
contribute primarily to the intercept at accessible times.  For deeper cylinder
partitions, it may also modify the effective slope because several transition
types and source-state occupations enter.  Therefore, the following relation
should be considered an ensemble-level finite-time ansatz rather than a
direct consequence of a pathwise limit theorem.

By considering the ensemble average in Eq.~\eqref{eq:IP_BLn_Gamma_form} and dividing
by \(\langle G_n\rangle_{\rm ens}\), we obtain
\begin{align}
    \frac{\langle I_P(n)\rangle_{\rm ens}}
    {\langle G_n\rangle_{\rm ens}}
    &=
    L_n
    \frac{\langle B_{P,n}\rangle_{\rm ens}}
    {\langle G_n\rangle_{\rm ens}}
    \nonumber\\
    &\quad+
    \frac{
    \left\langle
    B_{P,n}
    \left[
        \Gamma_P(n,x_0)
        -
        \log\frac{B_{P,n}}{a_n}
    \right]
    \right\rangle_{\rm ens}
    }
    {\langle G_n\rangle_{\rm ens}} .
    \label{eq:IPG_decomposition_before_eta}
\end{align}
The Hopf ratio argument in Eq.~\eqref{eq:Hopf_ratio_BP_G} suggests that
\(B_{P,n}\) and \(G_n\) are governed by the same return activity.  Thus, the
first ratio in Eq.~\eqref{eq:IPG_decomposition_before_eta} yields the basic
source of the \(L_n\)-linear term.  The second ratio contains finite-time and
partition-dependent corrections.  For the dyadic partition these corrections
are expected to contribute primarily to the intercept.  However, for deeper cylinder
partitions, the source-state occupation frequencies in
\(\Gamma_P\) may produce corrections that vary on the same finite-time scale
as \(L_n\).  These contributions are absorbed into the effective fitted slope,
whereas the remaining \(O(1)\)-type part is absorbed into the intercept.  This
motivates the ensemble-level finite-time ansatz
\begin{equation}
    \frac{\langle I_P(n)\rangle_{\rm ens}}
    {\langle G_n\rangle_{\rm ens}}
    \simeq
    \eta_{1,P}(B_{\rm MB})L_n+
    \eta_{2,P}(B_{\rm MB}),
    \label{eq:IP_G_linear_relation}
\end{equation}
or equivalently
\begin{equation}
    \langle I_P(n)\rangle_{\rm ens}
    \simeq
    \bigl[
    \eta_{1,P}(B_{\rm MB})L_n+
    \eta_{2,P}(B_{\rm MB})
    \bigr]
    \langle G_n\rangle_{\rm ens} .
    \label{eq:IP_G_linear_relation_equiv}
\end{equation}
Here, \(B_{\rm MB}\) denotes the parameter of the modified Bernoulli map,
whereas \(B_{P,n}\) denotes the transition count.  The coefficients
\(\eta_{1,P}(B_{\rm MB})\) and \(\eta_{2,P}(B_{\rm MB})\) are finite-time
effective coefficients, not universal entropy constants.  The slope
\(\eta_{1,P}\) contains the basic Hopf ratio contribution between
\(B_{P,n}\) and \(G_n\), and, for deeper partitions, may also absorb
\(L_n\)-scale finite-time corrections from the bracketed term in
Eq.~\eqref{eq:IP_BLn_Gamma_form}.  The intercept \(\eta_{2,P}\) collects the
remaining corrections that do not increase linearly with \(L_n\) over the fitting
window.  Thus, Eq.~\eqref{eq:IP_G_linear_relation} should be interpreted as a
partition-dependent finite-time relation and not as a theorem or Pesin-type identity.

\begin{figure*}
    \centering
    \includegraphics[width=0.90\textwidth]{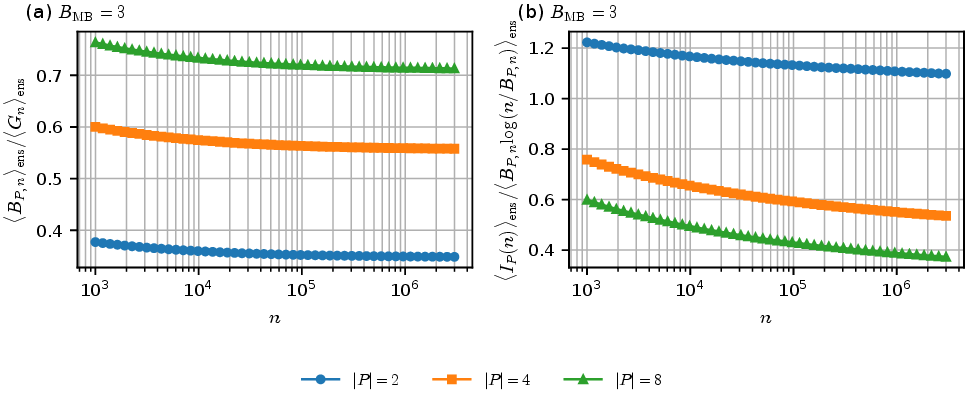}
    \caption{Sparse-transition diagnostics for the modified Bernoulli map with \(B_{\rm MB}=3\).
    (a) Ratio \(\langle B_{P,n}\rangle_{\rm ens}/\langle G_n\rangle_{\rm ens}\) for cylinder partitions \(|P|=2,4,8\).  The slow variation indicates that the non-self-transition count and the generalized Lyapunov sum are controlled by the same finite-time intermittent scale.
    (b) Ratio \(\langle I_P(n)\rangle_{\rm ens}/\langle B_{P,n}\log(n/B_{P,n})\rangle_{\rm ens}\).  For the dyadic partition \(|P|=2\), the sparse-transition Markov-information interpretation is most direct; for deeper cylinder partitions, the denominator uses only the total non-self-transition count and the ratio includes partition-dependent transition-type corrections.  Initial conditions were sampled independently from the Lebesgue distribution
on \((0,1)\); the ensemble size was \(M=10000\).  The maximum horizon was
\(n=3\times10^6\), where \(n\) counts the number of empirical
cylinder-state transitions for \(I_P(n)\) and \(B_{P,n}\) and the
number of terms in \(G_n=\sum_{k=0}^{n-1}\log|T'(x_k)|\).  For the largest
partition shown, \(|P|=8\) (depth \(m=3\)), constructing the cylinder
labels requires the underlying scalar orbit up to \(x_{n+2}\).}
    \label{fig:sec6_transition_count_sparse_coding}
\end{figure*}

We test this picture numerically for the modified Bernoulli map.
Figure~\ref{fig:sec6_transition_count_sparse_coding} shows two diagnostics
for \(B_{\rm MB}=3\).  Panel~(a) shows that, for cylinder partitions with
\(|P|=2,4,8\), the ratio
\begin{equation}
    \frac{\langle B_{P,n}\rangle_{\rm ens}}
    {\langle G_n\rangle_{\rm ens}}
    \label{eq:BPn_Gn_ratio_fig8a}
\end{equation}
varies slowly with \(n\), supporting the idea that \(B_{P,n}\) and \(G_n\)
are controlled by the same finite-time scale.  Panel~(b) shows the ratio
\begin{equation}
    \frac{\langle I_P(n)\rangle_{\rm ens}}
    {\langle B_{P,n}\log(n/B_{P,n})\rangle_{\rm ens}} .
    \label{eq:fig8b_ratio_def}
\end{equation}
For the dyadic partition, the non-self-transitions are only the two crossing
types \(X_1\to X_2\) and \(X_2\to X_1\).  In this case, the denominator in
Eq.~\eqref{eq:fig8b_ratio_def} matches the coarse-grained sparse-transition
self-information in Eq.~\eqref{eq:IP_BlogB_expansion}, and the remaining
terms are expected to provide slower finite-time corrections.  Thus, the ratio 
is expected to vary slowly and to be close to unity up to finite-time effects.  This is
consistent with the slowly varying plateau observed at \(|P|=2\).

For deeper cylinder partitions, the same ratio should be interpreted more
cautiously.  The denominator
\(\langle B_{P,n}\log(n/B_{P,n})\rangle_{\rm ens}\) uses only the total
non-self-transition count and therefore combines several transition types into
one coarse count.  The correction term \(\Gamma_P B_{P,n}\), particularly the
part involving the source-state occupation frequencies \(\rho_i^{(n)}\), can
be comparable to the leading logarithmic term at accessible times.  Thus, the ratios 
for \(|P|=4,8\) need not be close to unity.  Their slow variation still
indicates that \(B_{P,n}\log(n/B_{P,n})\) captures a relevant
sparse-transition scale, but the effective coefficients must be considered as
partition dependent.

\begin{figure*}
    \centering
    \includegraphics[width=0.90\textwidth]{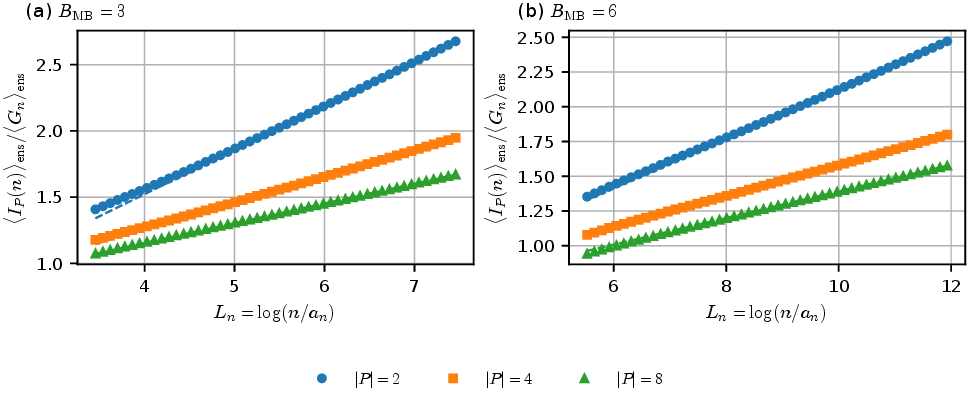}
    \caption{Linear-in-\(L_n\) relation between the empirical Markov information and generalized Lyapunov sums.  The vertical axis is \(\langle I_P(n)\rangle_{\rm ens}/\langle G_n\rangle_{\rm ens}\), and the horizontal axis is \(L_n=\log(n/a_n)\).  The dashed lines show linear fits with an intercept over the largest \(50\%\) of the logarithmically spaced checkpoints.  The observed dependence supports Eq.~\eqref{eq:IP_G_linear_relation}.  Initial conditions were sampled independently from the Lebesgue distribution
on \((0,1)\); the ensemble size was \(M=10000\).  The maximum horizon was
\(n=3\times10^6\), where \(n\) counts the number of empirical
cylinder-state transitions used for \(I_P(n)\) and the number of terms in
\(G_n\).  For the largest partition shown, \(|P|=8\) (depth \(m=3\)),
constructing the cylinder labels requires the underlying scalar orbit up to
\(x_{n+2}\).}
    \label{fig:sec6_linear_relation_gn}
\end{figure*}

Figure~\ref{fig:sec6_linear_relation_gn} directly tests
Eq.~\eqref{eq:IP_G_linear_relation}.  The vertical axis is
\(\langle I_P(n)\rangle_{\rm ens}/\langle G_n\rangle_{\rm ens}\), and
the horizontal axis is \(L_n=\log(n/a_n)\).  For both \(B_{\rm MB}=3\) and
\(B_{\rm MB}=6\), and for all displayed cylinder resolutions, the data are
well described over the tail window by a linear function of \(L_n\) with a
nonzero intercept.    This comparison remains
meaningful, even when \(\langle G_n/a_n\rangle_{\rm ens}\) still exhibits
finite-time drift, because \(G_n\) itself is used as the measured finite-time
scale.

\begin{figure*}
    \centering
    \includegraphics[width=0.90\textwidth]{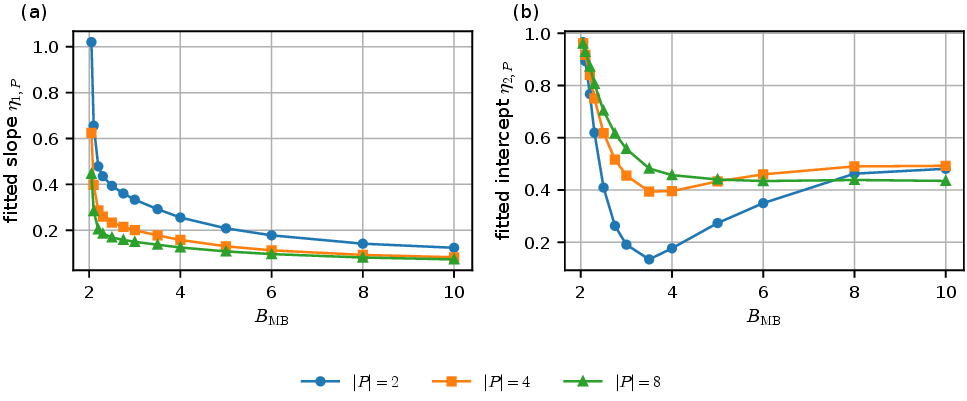}
    \caption{Fitted slope and intercept in the relation
    \(\langle I_P(n)\rangle_{\rm ens}/\langle G_n\rangle_{\rm ens}\simeq
    \eta_{1,P}(B_{\rm MB})L_n+\eta_{2,P}(B_{\rm MB})\).
    The coefficients were obtained by fitting Eq.~\eqref{eq:IP_G_linear_relation} over the largest \(50\%\) of the logarithmically spaced checkpoints.  Their dependence on both the map parameter \(B_{\rm MB}\) and the partition \(P\) indicates that the relation is partition dependent rather than a universal entropy identity.  For each \(B_{\rm MB}\), initial conditions were sampled independently from
the Lebesgue distribution on \((0,1)\); the ensemble size was \(M=10000\).
The maximum horizon was \(n=3\times10^6\), where \(n\) counts the
number of empirical cylinder-state transitions used for \(I_P(n)\) and the
number of terms in \(G_n\).  For the largest partition used in the fit,
\(|P|=8\) (depth \(m=3\)), constructing the cylinder labels requires the
underlying scalar orbit up to \(x_{n+2}\).}
    \label{fig:sec6_eta_coefficients}
\end{figure*}

Figure~\ref{fig:sec6_eta_coefficients} shows the fitted slope and
intercept in Eq.~\eqref{eq:IP_G_linear_relation} for a broader range of
\(B_{\rm MB}\).  Both coefficients depend on \(B_{\rm MB}\) and the
partition resolution.  This dependence is important because it shows that
Eq.~\eqref{eq:IP_G_linear_relation} is not a universal entropy identity.
Instead, it quantifies the representation of the intermittent activity measured by the
finite-time generalized Lyapunov sum, under the selected finite
symbolic observation, as a one-step empirical Markov information sum.

For the dyadic partition \(|P|=2\), the fitted slope exhibits a particularly
simple trend.  Over the parameter range in which the finite-time drift is moderate,
the numerical data suggest
\begin{equation}
    \eta_{1,P_2}(B_{\rm MB})
    \simeq
    \frac{1}{B_{\rm MB}}
    =
    \frac{\alpha}{1+\alpha},
    \qquad
    \alpha=\frac{1}{B_{\rm MB}-1}.
    \label{eq:eta1_dyadic_prediction}
\end{equation}
This observation can be understood from a more explicit escape estimate near
an indifferent fixed point.  First, we consider the left fixed point \(x=0\). The
right fixed point is treated in the same manner after replacing \(x\) with \(1-x\).
Near \(x=0\) the modified Bernoulli map has the local form
\begin{equation}
    T(x)-x \simeq A x^{B_{\rm MB}},
    \qquad A=2^{B_{\rm MB}-1}.
    \label{eq:local_MB_escape}
\end{equation}
During a laminar phase, the orbit moves slowly away from the indifferent
fixed point.  Replacing the discrete dynamics with continuous-time
approximation
\begin{equation}
    \frac{dx}{d\tau}=A x^{B_{\rm MB}},
    \label{eq:continuous_escape_approx}
\end{equation}
we obtain, for an escape from an initial point \(x_0\ll 1\) to a fixed cutoff
\(x_c=O(1)\),
\begin{equation}
    \tau(x_0)
    \simeq
    \int_{x_0}^{x_c}\frac{dx}{A x^{B_{\rm MB}}}
    =
    \frac{x_0^{1-B_{\rm MB}}-x_c^{1-B_{\rm MB}}}
    {A(B_{\rm MB}-1)} .
    \label{eq:escape_time_x0}
\end{equation}
Thus the laminar time has the usual power-law scaling
\begin{equation}
    \tau(x_0)\asymp x_0^{1-B_{\rm MB}},
    \qquad
    \alpha=\frac{1}{B_{\rm MB}-1}.
    \label{eq:laminar_time_scaling_sec6}
\end{equation}
This is the same continuous-time escape picture commonly used for
intermittent maps with marginally unstable fixed points
\cite{korabel2010separation}.

Next, we estimate the Lyapunov contribution accumulated during this escape.
Because
\begin{equation}
    T'(x)
    =
    1+A B_{\rm MB}x^{B_{\rm MB}-1},
    \label{eq:local_derivative_MB}
\end{equation}
for \(x\ll1\),we obtain
\begin{equation}
    \log |T'(x)|
    \simeq
    A B_{\rm MB}x^{B_{\rm MB}-1}.
    \label{eq:local_log_derivative_MB}
\end{equation}
Using \(d\tau=dx/(A x^{B_{\rm MB}})\) in the continuous approximation, we obtain the
Lyapunov contribution of one laminar escape as
\begin{align}
    \mathcal G(x_0)
    &\simeq
    \int_{x_0}^{x_c}
    A B_{\rm MB}x^{B_{\rm MB}-1}
    \frac{dx}{A x^{B_{\rm MB}}}
    \nonumber\\
    &=
    B_{\rm MB}\int_{x_0}^{x_c}\frac{dx}{x}
    =
    B_{\rm MB}\log\frac{x_c}{x_0}.
    \label{eq:single_escape_lyapunov}
\end{align}
Equation~\eqref{eq:escape_time_x0} yields
\begin{equation}
    \log\frac{x_c}{x_0}
    =
    \frac{1}{B_{\rm MB}-1}\log \tau
    +O(1)
    =
    \alpha\log\tau+O(1),
    \label{eq:log_xc_x0_tau}
\end{equation}
such that
\begin{equation}
    \mathcal G(x_0)
    \simeq
    B_{\rm MB}\alpha\log\tau+O(1).
    \label{eq:single_escape_lyapunov_tau}
\end{equation}
The laminar-time distribution has a tail of the form
\begin{equation}
    \psi(\tau)\sim C\tau^{-1-\alpha}.
    \label{eq:laminar_tail}
\end{equation}
Although the mean laminar time diverges, the logarithmic moment is finite:
\begin{equation}
    \int_1^\infty \alpha \tau^{-1-\alpha}\log\tau\,d\tau
    =
    \frac{1}{\alpha}.
    \label{eq:log_moment_tau}
\end{equation}
Consequently, the typical Lyapunov contribution of a completed laminar escape
is of the order
\begin{equation}
    B_{\rm MB}\alpha\times \frac{1}{\alpha}
    =
    B_{\rm MB},
    \label{eq:escape_lyapunov_order_B}
\end{equation}
up to constants depending on the cutoff, reinjection statistics, and the
non-laminar part of the excursion.  However, for the dyadic
partition \(P_2\), a completed escape from one laminar side to the other
increases the cross-transition count \(B_{P_2,n}\) by approximately one.
Therefore, if \(N_n\) denotes the number of completed laminar escapes up to
time \(n\), the heuristic estimates are
\begin{equation}
    B_{P_2,n}\approx N_n,
    \qquad
    G_n\approx B_{\rm MB}N_n.
    \label{eq:BP2_Gn_Nn_estimate}
\end{equation}
This gives
\begin{equation}
    \frac{B_{P_2,n}}{G_n}
    \approx
    \frac{1}{B_{\rm MB}}.
    \label{eq:dyadic_transition_per_G}
\end{equation}
Finally, the leading part of the dyadic empirical Markov information sum is
the rare-transition count multiplied by the rare-transition self-information factor,
\begin{equation}
    I_{P_2}(n)
    \simeq
    B_{P_2,n}L_n+O(B_{P_2,n}),
    \qquad
    L_n=\log\frac{n}{a_n}.
    \label{eq:IP2_BPn_Ln}
\end{equation}
Dividing by \(G_n\) and using Eq.~\eqref{eq:dyadic_transition_per_G}, we
obtain the leading prediction
\begin{equation}
    \frac{I_{P_2}(n)}{G_n}
    \simeq
    \frac{1}{B_{\rm MB}}L_n+O(1).
    \label{eq:IP2_over_Gn_prediction}
\end{equation}
This is the heuristic origin of Eq.~\eqref{eq:eta1_dyadic_prediction}.  The
argument is not meant to determine the intercept nor the finite-time
corrections, but it explains why the slope with respect to \(L_n\) should be
close to \(1/B_{\rm MB}\) for the dyadic partition.

\begin{figure*}
    \centering
    \includegraphics[width=0.90\textwidth]{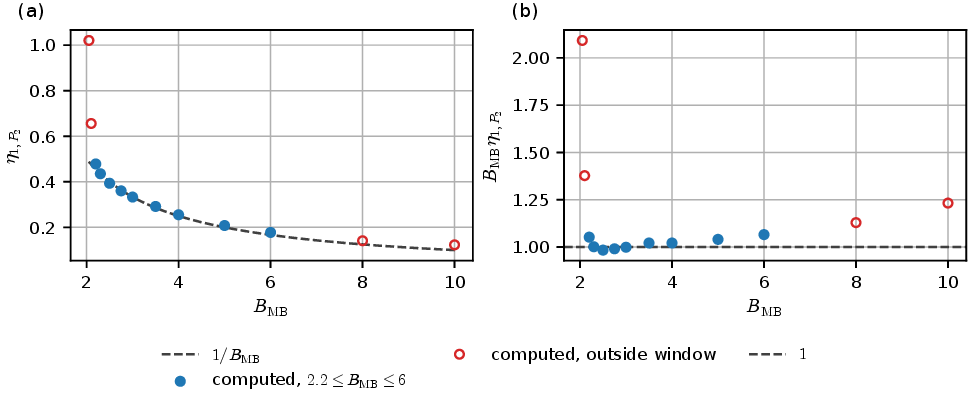}
    \caption{Dyadic-partition slope \(\eta_{1,P_2}\) compared with the heuristic prediction \(1/B_{\rm MB}=\alpha/(1+\alpha)\). The filled symbols indicate the parameter interval where the finite-time drift in the present simulations is moderate; the open symbols indicate values outside this interval.  Panel (b) shows the same data as \(B_{\rm MB}\eta_{1,P_2}\), for which the prediction becomes the horizontal line at one.  For each \(B_{\rm MB}\), initial conditions were sampled independently from
the Lebesgue distribution on \((0,1)\); the ensemble size was \(M=10000\).
The maximum horizon was \(n=3\times10^6\); for the dyadic partition
\(P_2\), this is the number of empirical dyadic transitions used for
\(I_{P_2}(n)\) and the number of terms in \(G_n\).}
    \label{fig:sec6_eta1_dyadic_prediction}
\end{figure*}

Figure~\ref{fig:sec6_eta1_dyadic_prediction} compares the fitted
\(\eta_{1,P_2}\) with \(1/B_{\rm MB}\).  Points outside the highlighted
interval, particularly close to \(B_{\rm MB}=2\) and for very large
\(B_{\rm MB}\), exhibit stronger finite-time deviations.  This is consistent
with the slow convergence of the generalized Lyapunov sums near the transition
point and for small \(\alpha\).  Therefore, we consider
Eq.~\eqref{eq:eta1_dyadic_prediction} as a finite-time numerical observation
supported by an escape-scaling argument, not as an additional universal
Pesin-type identity.

The conclusion of this section is twofold.  First, \(I_P(n)\) and \(G_n\) do
not have the same growth scale: \(G_n\) is a return-sequence-scale quantity,
whereas \(I_P(n)\) includes an additional logarithmic self-information contribution associated with rare transitions.  Second, the two quantities are related.
The transition count underlying \(I_P(n)\) is controlled using the same
finite-time intermittent scale as \(G_n\), and the ensemble mean of \(I_P(n)\)
is well described, relative to \(\langle G_n\rangle_{\rm ens}\), by a linear
function of \(L_n=\log(n/a_n)\) with an intercept.  Thus, the empirical
one-step conditional entropy does not provide another estimator of Krengel
entropy, but its information sum provides a computable, partition-dependent
measure of sparse-transition information in weakly chaotic intermittent
dynamics.

\section{Discussion}

We have studied the empirical one-step conditional entropy of a fixed finite partition in one-dimensional intermittent maps using an infinite invariant measure.  The central point of this study is that this quantity has two distinct aspects.  As a raw per-step entropy rate, it vanishes and, therefore, cannot serve as a nonzero infinite-time scalar measure of weak chaos.  However, as a finite-time empirical Markov information sum, it retains a structured sparse-transition scaling and nontrivial self-normalized fluctuations.

The first result is the vanishing theorem described in Sec.~III.  For the modified Bernoulli map and the Boole transformation, \(H_P^{(1,n)}(x_0)\to0\) for fixed partitions of the type considered here.  This is not a statement that the dynamics has no complexity.  Instead, in the infinite-measure regime, typical trajectories spend most of their time in long laminar phases near indifferent fixed points.  Therefore, at the level of the empirical one-step transition model induced by the fixed
partition, the transition matrix is nearly diagonal: Most observed
transitions are self-transitions with empirical probabilities close to one,
and the average amount of information per time step collapses to zero.

The second result is the sparse-transition scaling of the ensemble mean.  For the information sum \(I_P(n)=nH_P^{(1,n)}\), the rare transitions between laminar regions occur on the return-sequence scale \(a_n\).  Each such transition has one-step conditional self-information of the order
\(\log(n/a_n)\).  This leads to
\[
    \langle I_P(n)\rangle_{\rm ens}
    \simeq
    c_{1,P}a_n\log\frac{n}{a_n}
    +
    c_{2,P}a_n,
\]
or equivalently
\[
    \langle H_P^{(1,n)}\rangle_{\rm ens}
    \simeq
    c_{1,P}\frac{a_n}{n}\log\frac{n}{a_n}
    +
    c_{2,P}\frac{a_n}{n}.
\]
The \(O(a_n)\) correction is not merely cosmetic.  It is visible in the finite-time simulations, particularly near the transition between finite and infinite invariant measures, where \(\log(n/a_n)\) increases gradually.

The third result concerns fluctuations.  Although \(H_P^{(1,n)}\) tends to zero, the self-normalized variable
\[
    Z_P^{(n)}
    =
    \frac{H_P^{(1,n)}}{\langle H_P^{(1,n)}\rangle_{\rm ens}}
\]
has a nontrivial distribution.  The DKA theorem does not apply directly to \(I_P(n)\) because the one-step conditional self-information depends on the empirical transition
probabilities and, therefore, on the observation time \(n\).  Nevertheless, the sparse-transition picture suggests that the dominant fluctuations are governed by the same intermittent return process.  Numerical results using fixed cylinder partitions of moderate resolution are consistent with normalized Mittag--Leffler distributions.  The leading sparse-transition normalization \(I_P/(a_n\log(n/a_n))\) provides an \(O(1)\) mean up to \(O(1/\log(n/a_n))\) corrections, but the role of self-normalization is more basic: it removes the deterministic mean scale while preserving trajectory-to-trajectory weak-chaos fluctuations.

The fourth result clarifies the relationship to generalized Lyapunov sums and orbit complexity.  The generalized Lyapunov sum
\[
    G_n=\sum_{k=0}^{n-1}\log|T'(x_k)|
\]
is a Birkhoff sum of a fixed integrable observable and belongs to the return-sequence scale \(a_n\).  Lempel--Ziv complexity in infinite-measure systems is also naturally described through ratio-type relations with such return-sequence-scale sums.  In contrast, \(I_P(n)\) increases on the larger scale \(a_n\log(n/a_n)\) because it accumulates finite-memory empirical Markov self-information over rare transitions.  Thus, \(I_P(n)\) is not a Krengel entropy estimator and does not provide another generalized Pesin identity.

However, these two factors are not unrelated.  Our numerical results show that
\[
    \frac{\langle I_P(n)\rangle_{\rm ens}}
    {\langle G_n\rangle_{\rm ens}}
    \simeq
    \eta_{1,P}(B_{\rm MB})\log\frac{n}{a_n}
    +
    \eta_{2,P}(B_{\rm MB}).
\]
This relationship should be interpreted as a partition-dependent finite-time relationship between the empirical Markov information and generalized Lyapunov sums \(G_n\).  In the dyadic case, the fitted slope is close to \(1/B_{\rm MB}\) over the parameter range in which the finite-time corrections are moderate.  A simple escape estimate near the indifferent fixed points explains this trend: one completed laminar escape contributes \(O(B_{\rm MB})\) to the Lyapunov sum but only one dyadic crossing to the transition count.  Subsequently, the empirical information sum adds the sparse-transition
self-information factor \(\log(n/a_n)\).

Several limitations should be noted.  First, the results for the ensemble mean beyond the leading upper-bound argument are scaling ansatzes rather than full-limit theorems.  Second, the coefficients \(c_{1,P},c_{2,P},\eta_{1,P},\eta_{2,P}\) depend on the partition and finite-time details.  Therefore, they should not be treated as universal entropy invariants.  Third, the distributional comparison presented in Sec.~V depends on the use of fixed partitions with sufficient resolution.  Very coarse partitions, particularly the dyadic partition, exhibit stronger finite-time deviations because the information sum is dominated by a single crossing count.  Conversely, partitions that are too fine may suffer from finite-sample sparsity in the empirical transition matrix.  The simulations used fixed moderate cylinder partitions to balance these effects.

The map-blind equal-bin test in Appendix~\ref{app:equal-bin}
is relevant to time-series applications.  Generalized
Lyapunov sums require the derivative of the map, and
Krengel-entropy-based evaluations require information about
the infinite invariant density or an induced transformation.
Previous Lempel--Ziv complexity studies of weak chaos
used symbolic codings adapted to the underlying map and were aimed
at entropy--complexity relations such as generalized Pesin-type
identities. We do not analyze such algorithmic-complexity estimators
here. Instead, \(H_P^{(1,n)}\) is constructed from empirical state and
transition counts of a chosen finite partition and therefore retains an
explicit one-step transition-matrix interpretation.
The equal-bin results show that, at least for the benchmark
intermittent maps studied here, the self-normalized
Mittag--Leffler-type fluctuation persists even when the
symbolic observation is selected without using the explicit
map structure.

The practical implication is that the empirical one-step conditional entropy remains useful in weak chaos but not as a raw entropy rate.  Its value lies in revealing how finite symbolic observations encode rare transitions.  The raw entropy rate \(H_P^{(1,n)}\) detects the collapse of per-step unpredictability, \(I_P(n)\) quantifies the cumulative empirical Markov self-information of sparse transitions, \(Z_P^{(n)}\) captures nontrivial Mittag--Leffler-type fluctuations, and the comparison with \(G_n\) shows the relation of this self-information contribution to the same intermittent activity underlying generalized Lyapunov statistics.

Future research should aim to prove sharper limit theorems for \(I_P(n)\) as an \(n\)-dependent empirical observable, clarify the partition dependence of the coefficients, and extend the analysis to higher-order Markov approximations and experimentally observed time series.  Another important direction is to determine the change in the results when the partition resolution is allowed to vary with \(n\) because the present paper maintains the symbolic partition fixed throughout the asymptotic analysis. The equal-bin test should also be interpreted as a finite-time
robustness check and not as a theorem for arbitrary data-adaptive partitions.  As shown in the appendix, the partition is fixed
after the observed range has been determined; if the partition
were re-estimated for each trajectory or observation
time, the additional partition fluctuations would be mixed with
the dynamical fluctuations.

\appendix

\section{Map-blind equal-bin partitions}
\label{app:equal-bin}

The numerical comparisons presented in Sec.~V were performed with
fixed cylinder partitions.  In this appendix, we report a
complementary robustness check using partitions constructed
directly from scalar time-series data.  The objective is to
illustrate the operational time-series characteristics of
\(H_P^{(1,n)}\): when a finite symbolic observation is fixed,
the statistic is computed using only the empirical one-step
transition counts.

For a prescribed observation time \(n\) and ensemble size
\(M\), we first generated the scalar trajectories
\[
  \{x_k^{(m)}: 1\le m\le M,\;0\le k\le n\}.
\]
The observed range was determined once from the pooled
data,
\[
  x_{\min}=\min_{m,k} x_k^{(m)},\qquad
  x_{\max}=\max_{m,k} x_k^{(m)}.
\]
For each \(K\), the interval \([x_{\min},x_{\max}]\) was
divided into \(K\) equal bins.  This partition was fixed
for all the trajectories in the ensemble.  No information about
the derivative of the map, the infinite invariant density, or
a cylinder partition was used to construct the
symbolic process.  The map was used only to generate
the benchmark scalar trajectories.

\begin{figure*}[t]
  \centering
  \includegraphics[width=\linewidth]{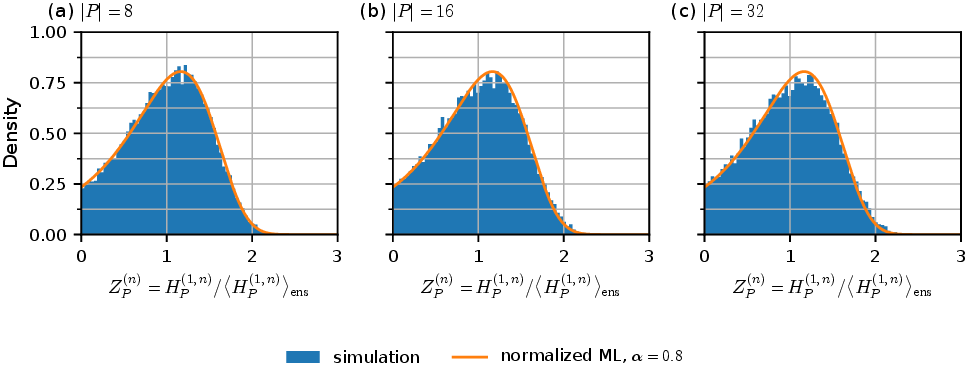}
  \caption{
  Map-blind equal-bin partition test for the modified Bernoulli
  map with \(B_{\rm MB}=2.25\), corresponding to \(\alpha=0.8\).
  The observed range \([x_{\min},x_{\max}]\) was determined once
  from the pooled scalar data
  \(\{x_k^{(m)}:1\le m\le M,0\le k\le n\}\) and then divided into
  \(K=8,16,32\) equal bins.  The partition was fixed for all
  trajectories.  The histograms show
  \(Z_P^{(n)}=H_P^{(1,n)}/\langle H_P^{(1,n)}\rangle_{\rm ens}\).
  The solid curve is the normalized Mittag--Leffler density with
  \(\alpha=0.8\).  The ensemble size was \(M=30000\), and \(n=10^5\) empirical one-step
transitions were counted.  Because the equal-bin symbol is determined directly
from \(x_k\), the required scalar orbit is \(x_0,\ldots,x_n\).
  }
  \label{fig:app_equalbin_mb}
\end{figure*}

\begin{figure*}[t]
  \centering
  \includegraphics[width=\linewidth]{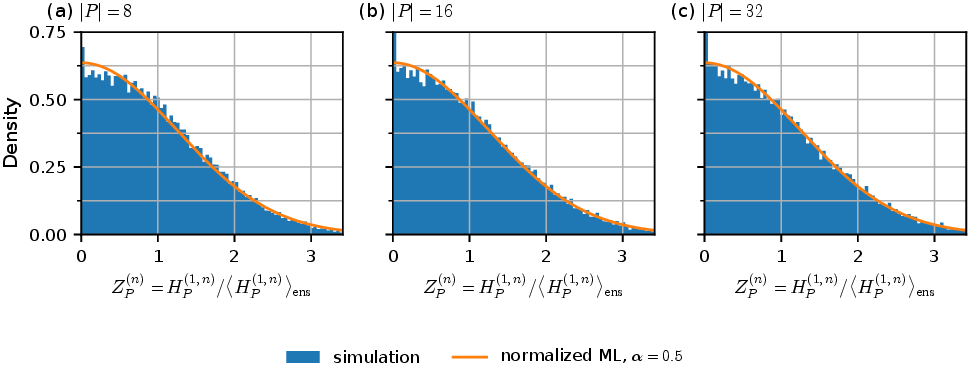}
  \caption{
  Map-blind equal-bin partition test for the Boole transformation, corresponding to $\alpha=1/2$.
  The observed range \([x_{\min},x_{\max}]\) was determined once
  from the pooled scalar data \(\{x_k^{(m)}:1\le m\le M,0\le k\le n\}\) and then divided into
  \(K=8,16,32\) equal bins.  The partition was fixed for all
  trajectories.  The histograms show
  \(Z_P^{(n)}=H_P^{(1,n)}/\langle H_P^{(1,n)}\rangle_{\rm ens}\).
  The solid curve is the normalized Mittag--Leffler density with
  \(\alpha=1/2\).  The ensemble size was \(M=30000\), and \(n=10^5\) empirical one-step
transitions were counted.  Because the equal-bin symbol is determined directly
from \(x_k\), the required scalar orbit is \(x_0,\ldots,x_n\).
  }
  \label{fig:app_equalbin_boole}
\end{figure*}

For the resulting symbolic sequences, we computed the same
empirical one-step conditional entropy \(H_P^{(1,n)}\) and
the self-normalized variable
\[
  Z_P^{(n)}
  =
  \frac{H_P^{(1,n)}}{\langle H_P^{(1,n)}\rangle_{\rm ens}}
  =
  \frac{I_P(n)}{\langle I_P(n)\rangle_{\rm ens}}.
\]
Figures~\ref{fig:app_equalbin_mb} and
\ref{fig:app_equalbin_boole} show that the histograms are
consistent with the corresponding normalized
Mittag--Leffler densities.  Thus, the self-normalized
Mittag--Leffler-type fluctuations observed in Sec.~V are not
specific to the cylinder partitions used in the main text.

This test should not be interpreted as a theorem for
arbitrary adaptive partitions.  The partition is fixed after
the pooled observed range has been determined.  If the range
or the bins were separately re-estimated for each trajectory
or each observation time, the observable itself would
become sample-dependent in an additional manner, and partition
fluctuations would be mixed with the dynamical fluctuations.

\begin{acknowledgments}
	The author is grateful to Prof. Takuma Akimoto, Dr. Soya Shinkai, and Dr. Yushi Nakano for their invaluable advice and informative discussions. This work was supported by JSPS KAKENHI Grant Numbers JP23K16963 and JP26K21338.
\end{acknowledgments}

\section*{AUTHOR DECLARATIONS}

\subsection*{Conflict of Interest}

The author declares no conflict of interest.

\subsection*{Author Contributions}

\textbf{Ken-ichi Okubo}: Conceptualization (equal); Methodology (equal); Formal analysis (equal); Investigation (equal);
Software (equal); Data curation (equal); Visualization (equal); Writing -- original draft (equal); Writing --
review and editing (equal); Funding acquisition (equal).

\section*{Data Availability}

The numerical data that support the findings of this study, including the data used to generate the figures, are available from the corresponding author upon reasonable request.

\bibliography{apssamp}

@article{ohya1998complexities,
  title={Complexities and their applications to characterization of chaos},
  author={Ohya, Masanori},
  journal={International Journal of Theoretical Physics},
  volume={37},
  number={1},
  pages={495--505},
  year={1998},
  publisher={Springer}
}

@article{inoue2000application,
  title={Application of chaos degree to some dynamical systems},
  author={Inoue, Kei and Ohya, Masanori and Sato, Keiko},
  journal={Chaos, Solitons \& Fractals},
  volume={11},
  number={9},
  pages={1377--1385},
  year={2000},
  publisher={Elsevier}
}

@article{mao2019investigation,
  title={Investigation of the difference between {Chaos} {Degree} and {Lyapunov} exponent for asymmetric tent maps},
  author={Mao, Tomoyuki and Okutomi, Hidetoshi and Umeno, Ken},
  journal={JSIAM Letters},
  volume={11},
  pages={61--64},
  year={2019},
  publisher={The Japan Society for Industrial and Applied Mathematics}
}

@article{inoue2021improved,
  title={An improved calculation formula of the extended entropic chaos degree and its application to two-dimensional chaotic maps},
  author={Inoue, Kei},
  journal={Entropy},
  volume={23},
  number={11},
  pages={1511},
  year={2021},
  publisher={MDPI}
}

@article{inoue2022analysis,
  title={Analysis of chaotic dynamics by the extended entropic {Chaos} degree},
  author={Inoue, Kei},
  journal={Entropy},
  volume={24},
  number={6},
  pages={827},
  year={2022},
  publisher={MDPI}
}

@article{adler1973ergodic,
  title={The ergodic infinite measure preserving transformation of {Boole}},
  author={Adler, Roy L and Weiss, Benjamin},
  journal={Israel Journal of Mathematics},
  volume={16},
  pages={263--278},
  year={1973},
  publisher={Springer}
}

@article{umeno2016exact,
  title={Exact {Lyapunov} exponents of the generalized Boole transformations},
  author={Umeno, Ken and Okubo, Ken-ichi},
  journal={Progress of Theoretical and Experimental Physics},
  volume={2016},
  number={2},
  pages={021A01},
  year={2016},
  publisher={Oxford University Press}
}

@article{okubo2018universality,
  title={Universality of the route to chaos: {Exact} analysis},
  author={Okubo, Ken-ichi and Umeno, Ken},
  journal={Progress of Theoretical and Experimental Physics},
  volume={2018},
  number={10},
  pages={103A01},
  year={2018},
  publisher={Oxford University Press}
}

@article{okubo2021infinite,
  title={Infinite ergodicity that preserves the {Lebesgue} measure},
  author={Okubo, Ken-ichi and Umeno, Ken},
  journal={Chaos: An Interdisciplinary Journal of Nonlinear Science},
  volume={31},
  number={3},
  pages={033135},
  year={2021},
  publisher={AIP Publishing}
}

@article{okubo2022universal,
  title={Universal critical behavior of transition to chaos: {Intermittency} route},
  author={Okubo, Ken-ichi and Umeno, Ken},
  journal={Progress of Theoretical and Experimental Physics},
  volume={2022},
  number={7},
  pages={073A01},
  year={2022},
  publisher={Oxford University Press}
}

@article{aizawa1989stagnant,
  title={Stagnant motions in {Hamiltonian} systems},
  author={Aizawa, Yoji and Kikuchi, Yasuhiro and Harayama, Takahisa and Yamamoto, Kenshi and Ota, Motonori and Tanaka, Kenji},
  journal={Progress of Theoretical Physics Supplement},
  volume={98},
  pages={36--82},
  year={1989},
  publisher={Oxford Academic}
}

@article{akimoto2007new,
  title={New aspects of the correlation functions in non-hyperbolic chaotic systems},
  author={Akimoto, Takuma and Aizawa, Yoji},
  journal={Journal of the Korean Physical Society},
  volume={50},
  number={1},
  pages={254},
  year={2007},
  publisher={Citeseer}
}

@article{akimoto2005large,
  title={Large fluctuations in the stationary-nonstationary chaos transition},
  author={Akimoto, Takuma and Aizawa, Yoji},
  journal={Progress of theoretical physics},
  volume={114},
  number={4},
  pages={737--748},
  year={2005},
  publisher={Oxford University Press}
}

@article{akimoto2003logarithmic,
  title={Logarithmic scaling in the stationary-nonstationary chaos transition},
  author={Akimoto, Takuma and Aizawa, Yoji},
  journal={Progress of theoretical physics},
  volume={110},
  number={5},
  pages={849--860},
  year={2003},
  publisher={Oxford University Press}
}

@book{aaronson1997introduction,
  title={An introduction to infinite ergodic theory},
  author={Aaronson, Jon},
  number={50},
  year={1997},
  publisher={American Mathematical Soc.}
}

@article{akimoto2010subexponential,
  title={Subexponential instability in one-dimensional maps implies infinite invariant measure},
  author={Akimoto, Takuma and Aizawa, Yoji},
  journal={Chaos: An Interdisciplinary Journal of Nonlinear Science},
  volume={20},
  number={3},
  pages={033110},
  year={2010},
  publisher={AIP Publishing}
}

@article{gaspard1988sporadicity,
  title={Sporadicity: between periodic and chaotic dynamical behaviors},
  author={Gaspard, Pierre and Wang, X-J},
  journal={Proceedings of the National Academy of Sciences},
  volume={85},
  number={13},
  pages={4591--4595},
  year={1988},
  publisher={National Acad Sciences}
}

@article{thaler2002limit,
  title={A limit theorem for sojourns near indifferent fixed points of one-dimensional maps},
  author={Thaler, Maximilian},
  journal={Ergodic Theory and Dynamical Systems},
  volume={22},
  number={4},
  pages={1289--1312},
  year={2002},
  publisher={Cambridge University Press}
}

@article{prykarpatski2024two,
  title={The two-dimensional boole-type transform and its ergodicity},
  author={Prykarpatski, AK and Balinsky, AA},
  journal={Journal of Mathematical Sciences},
  volume={278},
  number={6},
  pages={1055--1076},
  year={2024},
  publisher={Springer}
}

@incollection{klages2013weak,
  title={Weak chaos, infinite ergodic theory, and anomalous dynamics},
  author={Klages, Rainer},
  booktitle={From Hamiltonian Chaos to Complex Systems: A Nonlinear Physics Approach},
  pages={3--42},
  year={2013},
  publisher={Springer}
}

@article{korabel2009pesin,
  title={Pesin-type identity for intermittent dynamics with a zero {Lyapunov} exponent},
  author={Korabel, Nickolay and Barkai, Eli},
  journal={Physical Review Letters},
  volume={102},
  number={5},
  pages={050601},
  year={2009},
  publisher={APS}
}

@article{korabel2010separation,
  title={Separation of trajectories and its relation to entropy for intermittent systems with a zero {Lyapunov} exponent},
  author={Korabel, Nickolay and Barkai, Eli},
  journal={Physical Review E—Statistical, Nonlinear, and Soft Matter Physics},
  volume={82},
  number={1},
  pages={016209},
  year={2010},
  publisher={APS}
}

@article{inoue2023quantification,
  title={Quantification of chaos in a time series generated from a traffic flow model using the extended entropic chaos degree},
  author={Inoue, Kei and Tani, Kazuki},
  journal={Chaos, Solitons \& Fractals},
  volume={176},
  pages={114150},
  year={2023},
  publisher={Elsevier}
}

@article{bel2006weak,
  title={Weak ergodicity breaking with deterministic dynamics},
  author={Bel, Golan and Barkai, Eli},
  journal={Europhysics Letters},
  volume={74},
  number={1},
  pages={15},
  year={2006},
  publisher={IOP Publishing}
}

@article{geisel1984anomalous,
  title={Anomalous diffusion in intermittent chaotic systems},
  author={Geisel, Theo and Thomae, Stefan},
  journal={Physical Review Letters},
  volume={52},
  number={22},
  pages={1936},
  year={1984},
  publisher={APS}
}

@article{geisel1985accelerated,
  title={Accelerated diffusion in {Josephson} junctions and related chaotic systems},
  author={Geisel, T and Nierwetberg, J and Zacherl, A},
  journal={Physical Review Letters},
  volume={54},
  number={7},
  pages={616},
  year={1985},
  publisher={APS}
}

@article{barkai2003aging,
  title={Aging in subdiffusion generated by a deterministic dynamical system},
  author={Barkai, Eli},
  journal={Physical Review Letters},
  volume={90},
  number={10},
  pages={104101},
  year={2003},
  publisher={APS}
}

@article{barkai2021transitions,
  title={Transitions in the ergodicity of subrecoil-laser-cooled gases},
  author={Barkai, Eli and Radons, G{\"u}nter and Akimoto, Takuma},
  journal={Physical Review Letters},
  volume={127},
  number={14},
  pages={140605},
  year={2021},
  publisher={APS}
}

@article{barkai2022gas,
  title={Gas of sub-recoiled laser cooled atoms described by infinite ergodic theory},
  author={Barkai, Eli and Radons, G{\"u}nter and Akimoto, Takuma},
  journal={The Journal of Chemical Physics},
  volume={156},
  number={4},
  year={2022},
  pages={044118},
  publisher={AIP Publishing}
}

@article{akimoto2013aging,
  title={Aging generates regular motions in weakly chaotic systems},
  author={Akimoto, Takuma and Barkai, Eli},
  journal={Physical Review E—Statistical, Nonlinear, and Soft Matter Physics},
  volume={87},
  number={3},
  pages={032915},
  year={2013},
  publisher={APS}
}

@article{shinkai2006lempel,
  title={The {Lempel-Ziv} complexity of non-stationary chaos in infinite ergodic cases},
  author={Shinkai, Soya and Aizawa, Yoji},
  journal={Progress of theoretical physics},
  volume={116},
  number={3},
  pages={503--515},
  year={2006},
  publisher={Oxford University Press}
}

@article{shinkai2007lempel,
  title={The lempel-ziv complexity in infinite ergodic systems},
  author={Shinkai, Soya and Aizawa, Yoji},
  journal={Journal of the Korean Physical Society},
  volume={50},
  number={1},
  pages={261},
  year={2007},
  publisher={Citeseer}
}

@article{akimoto2015generalized,
  title={Generalized {Lyapunov} exponent as a unified characterization of dynamical instabilities},
  author={Akimoto, Takuma and Nakagawa, Masaki and Shinkai, Soya and Aizawa, Yoji},
  journal={Physical Review E—Statistical, Nonlinear, and Soft Matter Physics},
  volume={91},
  number={1},
  pages={012926},
  year={2015},
  publisher={APS}
}

@book{zaslavsky2007physics,
  title={The physics of chaos in {Hamiltonian} systems},
  author={Zaslavsky, George M},
  year={2007},
  publisher={World Scientific}
}

@article{lai2003noise,
  title={Noise-induced unstable dimension variability and transition to chaos in random dynamical systems},
  author={Lai, Ying-Cheng and Liu, Zonghua and Billings, Lora and Schwartz, Ira B},
  journal={Physical Review E—Statistical, Nonlinear, and Soft Matter Physics},
  volume={67},
  number={2},
  pages={026210},
  year={2003},
  publisher={APS}
}

@article{grebogi1984strange,
  title={Strange attractors that are not chaotic},
  author={Grebogi, Celso and Ott, Edward and Pelikan, Steven and Yorke, James A},
  journal={Physica D: Nonlinear Phenomena},
  volume={13},
  number={1-2},
  pages={261--268},
  year={1984},
  publisher={Elsevier}
}

@article{wang2004strange,
  title={Strange nonchaotic attractors in random dynamical systems},
  author={Wang, Xingang and Zhan, Meng and Lai, Choy-Heng and Lai, Ying-Cheng},
  journal={Physical Review Letters},
  volume={92},
  number={7},
  pages={074102},
  year={2004},
  publisher={APS}
}

@article{pomeau1980intermittent,
  title={Intermittent transition to turbulence in dissipative dynamical systems},
  author={Pomeau, Yves and Manneville, Paul},
  journal={Communications in Mathematical Physics},
  volume={74},
  pages={189--197},
  year={1980},
  publisher={Springer}
}

@article{lesne2014shannon,
  title={Shannon entropy: a rigorous notion at the crossroads between probability, information theory, dynamical systems and statistical physics},
  author={Lesne, Annick},
  journal={Mathematical Structures in Computer Science},
  volume={24},
  number={3},
  pages={e240311},
  year={2014},
  publisher={Cambridge University Press}
}

@book{arnold1968ergodic,
  author    = {Arnold, Vladimir Igorevich and Avez, Andr{\'e}},
  title     = {Ergodic problems of classical mechanics},
  publisher = {Benjamin, New York},
  year      = {1968},
}

@article{gaspard1993noise,
  title={Noise, chaos, and ($\varepsilon$, $\tau$)-entropy per unit time},
  author={Gaspard, Pierre and Wang, Xiao-Jing},
  journal={Physics reports},
  volume={235},
  number={6},
  pages={291--343},
  year={1993},
  publisher={Elsevier}
}

@article{cohen1985computing,
  title={Computing the Kolmogorov entropy from time signals of dissipative and conservative dynamical systems},
  author={Cohen, Aviad and Procaccia, Itamar},
  journal={Physical review A},
  volume={31},
  number={3},
  pages={1872},
  year={1985},
  publisher={APS}
}

@article{mao2026chaotic,
  title={Chaotic fluctuations mark the sign of mental activity in task-based heart rate variability},
  author={Mao, Tomoyuki and Okutomi, Hidetoshi and Umeno, Ken},
  journal={Scientific Reports},
  volume={16},
  number={1},
  pages={9221},
  year={2026},
  publisher={Nature Publishing Group UK London}
}

\end{document}